\documentclass[preprint,12pt,authoryear]{elsarticle}

\usepackage{amssymb}
\usepackage{amsmath}
\usepackage{color}
\usepackage{comment}
\usepackage{rotating}
\usepackage{multirow}
\usepackage{makecell}
\usepackage{tablefootnote}
\usepackage{epstopdf}
\usepackage{enumitem}
\usepackage{subcaption}
\usepackage{diagbox}
\usepackage{url}
\usepackage{graphicx}
\usepackage{threeparttable}
\usepackage{hyperref}
\usepackage{pdflscape}
\usepackage{ragged2e}
\usepackage{array}

\newcounter{excep}
\begin{document}

\begin{frontmatter}
\title{Price Distortions in Korea's Electricity Market: Barriers to Renewable Integration and Reform Pathways}

\author[inst1]{Ryungyeong Lee}
\author[inst2]{Minhan Yoon}
\author[inst1]{Hunyoung Shin}

\affiliation[inst1]{organization={Department of Electronic and Electrical Engineering},
            addressline={Hongik University}, 
            city={Seoul},
            postcode={04066}, 
            country={South Korea}}

\affiliation[inst2]{organization={Department of Electrical Engineering}
,
            addressline={Kwangwoon University}, 
            city={Seoul},
            postcode={01897}, 
            country={South Korea}}            

\begin{abstract}
Structural distortions in price signals within the Korean electricity market, governed by a cost-based pool (CBP) and a uniform pricing mechanism, fundamentally undermine the nation's energy transition goals. The current market design fails to reflect transmission constraints, real-time supply and demand dynamics, and generator-specific costs, leading to inefficient resource allocation and hindering long-term investments in renewable energy and grid flexibility. This paper identifies the key drivers of these distortions and proposes a holistic reform package to enhance market efficiency. The package includes four key reforms: \stepcounter{excep}(\roman{excep}) introducing a locational marginal pricing system to manage transmission constraints; \stepcounter{excep}(\roman{excep}) establishing a real-time market to reflect temporal value; \stepcounter{excep}(\roman{excep}) integrating market and system operations to resolve inconsistencies; and \stepcounter{excep}(\roman{excep}) transitioning from CBP to a price-based bidding system. Each reform targets a distinct source of inefficiency. The broader contribution of this study, however, lies in showing that, under the current Korean market design, the market cannot readily provide effective price signals. These reforms therefore need to be implemented jointly to establish a coherent market design in which price signals are aligned with Korea's energy policy objectives. 

\setcounter{excep}{0}
\end{abstract}



\begin{keyword}
Market Design \sep Price formation \sep Spot Market Reform \sep Korean Electricity Market


\end{keyword}

\end{frontmatter}


\section{Introduction} \label{Sec:Introduction}
The global transition to carbon neutrality presents significant challenges for many countries, especially those with high electricity demand. Korea, ranked seventh worldwide in electricity consumption with 577\,TWh \citep{EIA}, also aims to decarbonize its power sector to achieve carbon neutrality. To this end, the government has committed to reducing greenhouse gas emissions by 40\% from 2018 levels by 2030, a target specified in the Nationally Determined Contribution (NDC). They have attempted to significantly expand renewable energy to achieve this goal, as mandated by the national long-term energy roadmap. This plan, which is known as the Basic Plan for Long-term Electricity Supply and Demand, includes demand forecasts, capacity expansion strategies, and generation mix targets. In the 11th Basic Plan, published in 2025, renewable energy capacity is projected to reach 121.9\,GW by 2038, constituting 47.3\% of total installed capacity. 

However, a significant gap remains between these policy ambitions and the operational reality\textemdash renewable energy accounts for only about 10\% of generation in Korea. Although all power systems face complex hurdles in the transition to renewable energy \citep{SINSEL20202271, 9579026, BROWN2025107484}, the process in Korea faces specific systemic limits that hinder further progress. Compared to other countries, Korea lacks a fundamentally efficient price formation mechanism, which limits the efficient integration of variable renewable resources. 

As widely acknowledged, efficient price signals play a central role in electricity market performance. Hogan emphasizes that well-designed short-term price signals support efficient long-term market outcomes \citep{antonopoulos2020nodal}. \citep{HERRERO201542} similarly argues that long-term efficiency depends on the market price formation mechanism used in the wholesale market, which determines the quality of investment incentives. In other words, wholesale market prices formed particularly in the real-time market serve as efficient signals for short-term operations and long-term decisions for both suppliers and consumers \citep{FACCHINI2019110}. 

Despite this broad consensus, the Korean price formation mechanism exhibits non-granular price signals that fail to reflect physical grid constraints, such as transmission congestion and rapid supply--demand imbalances. The underlying cause is the rigid design of the Korean electricity market, which has remained unchanged since its inception\textemdash limited to a day-ahead market, a cost-based pool, and uniform price across mainland. Unfortunately, the current system delivers distorted price signals, which result in resource misallocation and operational inefficiencies across the entire power sector. 

Thus, this paper investigates why price formation fails in the Korean electricity market. We examine the design of the existing wholesale market and identifies three central deficiencies: absence of network constraints from price formation, discrepancies between market-clearing prices and actual dispatch outcomes, and inaccurate generator cost. Through theoretical analysis and bulk-system simulations, this work demonstrates how these institutional features distort market price signals. Based on this diagnosis, we propose four market design reforms to restore the functioning of price signals: the adoption of locational pricing, the introduction of a real-time market, the alignment of market clearing and dispatch through EMS (Energy Management System) enhancement, and the transition from the cost-based pool to a price-bidding mechanism.

The remainder of this paper is organized as follows. Section \ref{Sec2:Review} reviews the literature on electricity price formation and market redesign. Section \ref{Sec:Background} provides an overview of the Korean electricity sector, including its market structure, price formation mechanism, the evolution of market reforms, and the remaining challenges. Section \ref{Sec3:Diagnosis} examines the market failures that hinder renewable energy integration in the Korean electricity system, focusing on the geographic and operational characteristics of the system and the shortcomings of the current market design. Section \ref{Sec4:Package} proposes market design reforms aimed at addressing the identified issues and restoring effective price signals, presenting both a reform package and an implementation pathway. Section \ref{Sec5:Discussion} discusses the limitations of the proposed reforms and outlines additional considerations for future electricity market design.

\section{Literature Review} \label{Sec2:Review}
\subsection{Theoretical Foundations of Price Formation}
A primary objective of electricity market design is to provide efficient economic incentives for both short-term operation and long-term investment \citep{HOGAN201423}. The principles of price formation in electricity markets have been extensively studied over the past decades \citep{schweppe2013spot, HOGAN201423, cadwalader2010extended}. A fundamental principle, as emphasized by \citep{hogan1992contract, pope2014price}, is that market prices must align with the actual generation dispatch. This alignment implies that the market should be designed to ensure incentive compatibility, meaning that each participant can maximize their profit by adhering to the dispatch schedule determined by the system operator; in a centralized system where the operator manages both grid and market functions, this schedule is derived through unit commitment and security-constrained economic dispatch. Consequently, generating price signals consistent with actual dispatch enables the market to reach a competitive equilibrium that maximizes social welfare. In other words, ideal market prices reflect the true marginal value of electricity. Thus, the effectiveness of the pricing mechanism hinges on the precise definition of this marginal value. 

\subsection{Evolution of Electricity Market Design in a Changing Environment}
To adapt to the ongoing changes in the electricity sector, numerous studies emphasize the importance of market redesign to enhance market efficiency \citep{cramton2017electricity, CONEJO2018520, NEWBERY2018695, LYNCH2021101312}. Notably, \citep{NEWBERY2018695} argued that the electricity market should provide more granular and differentiated price signals. \citep{LESLIE2020106847} noted that the increasing deployment of renewable energy intensifies the need for accurate pricing mechanisms. In particular, \citep{LESLIE2025107489} explored this challenge within the Australian national electricity market, demonstrating that the misalignment between the wholesale market design and the physical realities of the power system, such as transmission congestion and losses, leads to distorted participant incentives and operational inefficiencies. 

In addition, many studies explore how to adapt market design to meet national policy objectives. \citep{MUNOZ2021111997} analyzed limitations in the Chilean wholesale electricity market that hinder the transition to low-carbon resources, focusing on solutions to improve price signals and investment incentives. Key issues they identified include the market's reliance on a merit-order curve (instead of dual variables) for pricing and the low temporal granularity of spot market prices. Examining China's power sector, \citep{LI2019330} analyzed how its characteristics constrain renewable energy integration and evaluated the effectiveness of various market-based approaches implemented across different regions. They concluded that while these approaches offer valuable lessons, effective market signals remain essential for promoting large-scale energy investment and integration. Furthermore, \citep{BROWN2025107484} argued that adopting locational marginal pricing is imperative for the effective integration of growing renewable resources into the Alberta wholesale market. 
 
\section{Institutional Background: Korean Market Design and Price Formation} \label{Sec:Background}
\subsection{Korean Wholesale Market Structure} 
Korea operates a centralized mandatory pool that has remained largely unchanged since its establishment in 2001. Governance is based on a strict functional separation. The Korea Power Exchange (KPX) serves as the independent system and market operator (ISO/MO), while the Korea Electric Power Corporation (KEPCO), a state-owned utility, retains a monopoly on transmission, distribution, and retail. 

\begin{figure}
    \centering
	\includegraphics[scale = 0.31]{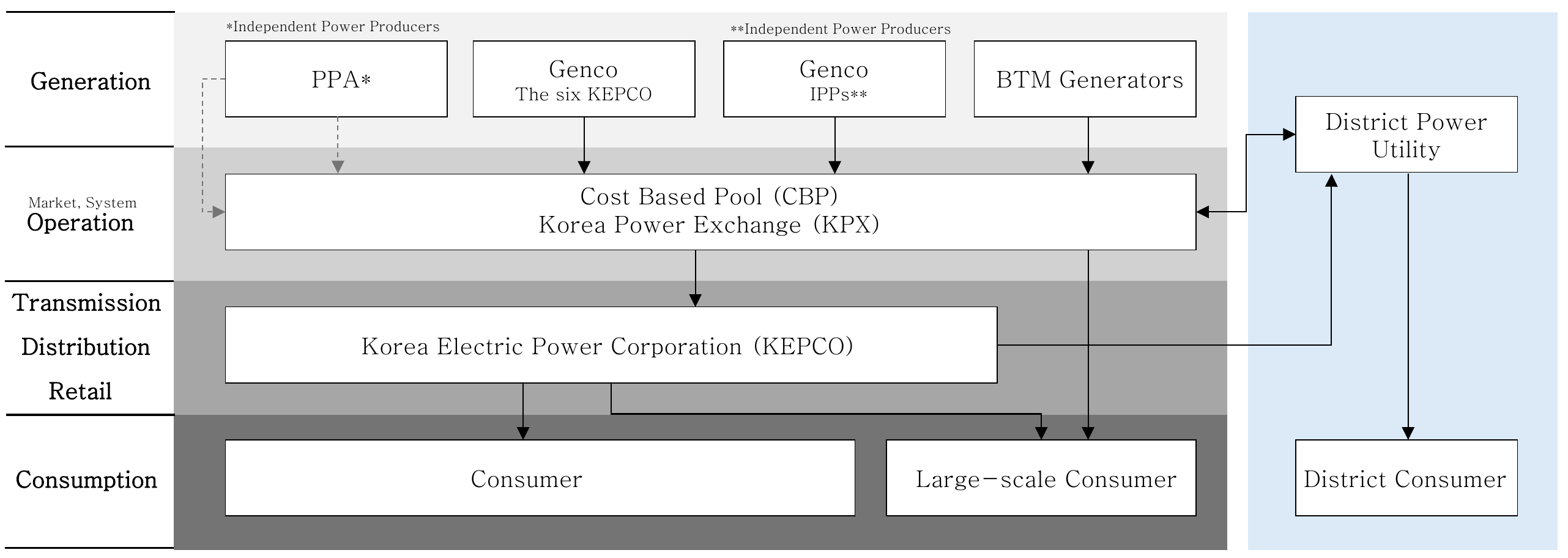}
	\caption{Korean Electricity Market Governance Structure}
	\label{fig:Korean Governance}
\end{figure}

Competition is effectively limited to the generation sector, which consists of KEPCO's six subsidiary generation companies (GenCos) and independent power producers (IPPs). However, the KEPCO GenCos remain dominant, accounting for approximately 55\% of the total installed capacity \citep{KPX2025}. All electricity trading must occur through the wholesale market administered by KPX, with KEPCO acting as the single buyer. Some exceptions to the mandatory pool exist for specific renewable transactions, but their trading volume remains minimal\footnote{Exceptions include: (i) small-scale renewable generators (less than 1 MW) contracting bilaterally with KEPCO; (ii) direct Power Purchase Agreements (PPAs) for renewable energy consumers; and (iii) community electricity providers. These do not materially change the centralized market structure.}. The overall governance of the Korean electricity market is summarized in Fig.~\ref{fig:Korean Governance}.

\begin{figure}
    \centering
    \includegraphics[scale = 0.4]{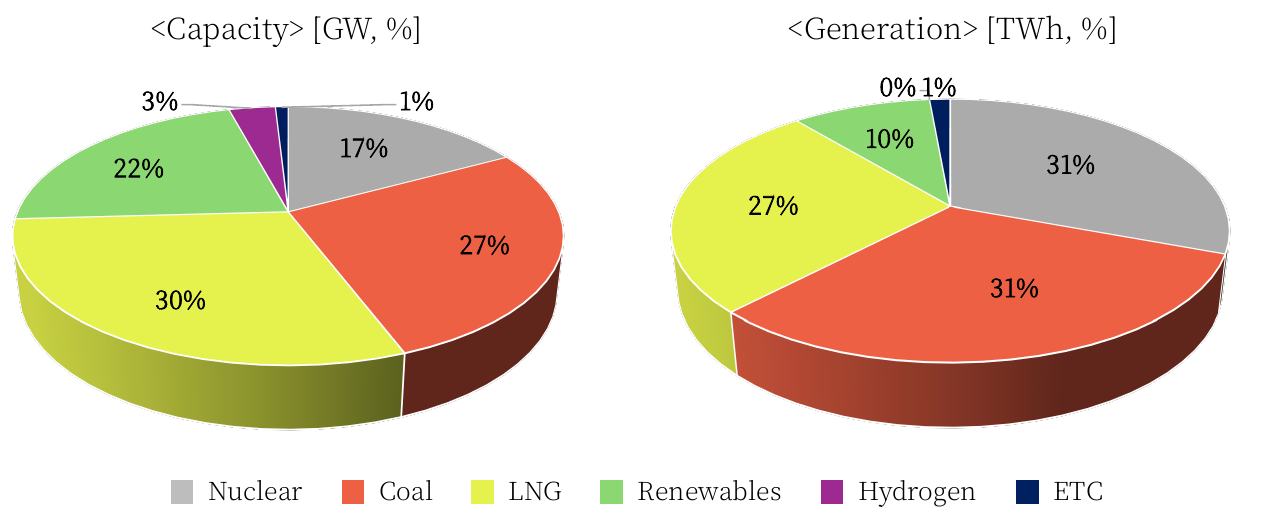}
    \caption{Korean Generation Mix (Left: Generator Capacity, Right: Generation Output)}
   \label{fig:Korea Mix}
\end{figure} 

As shown in Fig.~\ref{fig:Korea Mix}, the Korean generation mix is dominated by relatively inflexible baseload resources. According to the 11th Basic Plan for Electricity Supply and Demand, as of 2023, nuclear and coal power combined account for approximately 62\% of total generation, forming the backbone of supply. In contrast, renewable energy sources constitute only about 10\% of the generation mix.

\subsection{Price Formation Under the Cost-based Pool} \label{sec: price formation}
The System Marginal Price (SMP) serves as the wholesale electricity price in Korea. Unlike standard gross pools, where prices are determined through competitive bidding, the SMP is administratively calculated under a CBP mechanism. In the CBP, generators do not submit the price component of their offer curves; instead, KPX periodically assesses its cost parameters. Based on these assessed costs, KPX constructs a Day-Ahead Unit Commitment (DAUC) schedule that minimizes total generation cost subject to meeting forecasted demand and satisfying network constraints, stability constraints\footnote{It mainly refer to constraints on generator-side output and commitment used to maintain system stability, discussed in further detail in Section\ref{subsubsec:alignment}} and generator physical constraints such as ramp rates, minimum up/down times. 

Notably, the SMP is determined by the screened generator with the highest avoided cost based on the dispatch generation in the UC, rather than the dual variable of power balance constraints. The avoided cost of each generator is referred to as the Stack Price (SP), which for generator $i$ at hour $t$ is calculated reflecting its avoided cost: 
\begin{align}
    SP_{i,t} = IC_{i,t} + NLC_i / Q_{i,t} + SUC_i / (Q_{i,t} \times HoursOn_{i,t}) \label{eq: SP}
\end{align}
where $IC_{i,t}$ is the incremental cost including fuel cost, $NLC_i$ is the no-load cost, $SUC_i$ is start-up cost, $Q_{i,t}$ is the generation output, and $HoursOn_{i,t}$ is the duration the generator $i$ remains online at time $t$.

The SMP for each hour $t$ is then set by the highest SP among the selected marginal generators ($MG_t$):
\begin{align}
    SMP_{t} = \max\{SP_{i,t} ~|~ i \in MG_t\}. \label{eq: SMP}
\end{align}
Finally, the settlement revenue for generator $i$ is calculated by applying this day-ahead price to real-time output:
\begin{align}
    REV_{i,t} = SMP_{t} \times Q^{RT}_{i,t} \label{eq: RT settlement}
\end{align}
where $Q^{RT}_{i,t}$ denotes the actual real-time dispatch of generator $i$ at hour $t$.

This settlement rule results in three rigidities that obscure market signals:
\begin{itemize}
    \item \textit{Temporal Misalignment (Day-Ahead vs. Real-Time):} As shown in Eq. (\ref{eq: RT settlement}), prices determined hourly by the day-ahead ($SMP_t$) are applied to real-time quantities ($Q^{RT}$). When real-time conditions deviate from the day-ahead forecast, there is no market mechanism to value the flexibility required to close the gap. Instead, out-of-market payments (e.g., Make-Whole Payments (MWP), Day-Ahead Margin Assurance Payment (DAMAP) are used to compensate for these deviations administratively, rendering real-time scarcity value invisible.
    \item \textit{Spatial Uniformity:} Since the SMP in Eq. (\ref{eq: SMP}) is applied as a uniform price across mainland, the economic value of transmission constraints is not reflected in market price. Consequently, generators are compensated identically regardless of whether their generation relieves or exacerbates congestion, and as a result, the locational value of resources does not emerge as a price signal. 
    \item \textit{Administrative Valuation:} The generator cost parameters ($IC_{i,t}$, $SUC_i$, and $NLC_i$) to determine the SMP in Eq. (\ref{eq: SP}) are evaluated by the KPX. Because generators do not submit their own price offers, the resulting price cannot capture true costs that only the generator can observe such as fuel procurement costs, individual variable O\&M cost, or the scarcity value of capacity during peak periods.
\end{itemize}

\subsection{The 2022 Reform and Its Limitations}
\subsubsection{Pre-2022: Separated Pricing and Operations}
Prior to September 2022, the Korean electricity market operated with two parallel UC processes. One was a price-setting UC used to determine the SMP. That process followed an unconstrained merit-order dispatch, assuming a copper-plate network where the entire country was modeled as a single node without transmission constraints, thermal, and security constraints. The other was an operational scheduling UC used to determine the dispatch schedule, conducted under various system constraints such as transmission limits and N-1 reliability criteria. This gap resulted in out-of-market uplift payments \citep{__a1___2021}. These payments, known as Constraint-On (CON) and Constraint-Off (COFF) payments, compensate generators whose actual dispatch deviates from their price-setting schedule. As shown in Table \ref{tab:CON/COFF}, the total annual CON and COFF payments surged from KRW 795.5 billion to KRW 1,248.1 billion between 2013 and 2019. The rapid increase in these payments ultimately drove the 2022 market reform. 

\begin{table}[ht]
    \centering
    \caption{Annual payment of CON and COFF (hundred million KRW) \citep{__a1___2021}}
    \renewcommand{\arraystretch}{1.2}
    \begin{tabular}{c|c|c|c|c|c|c|c}
        \hline
        \hline
         & 2013 & 2014 & 2015 & 2016 & 2017 & 2018 & 2019 \\ 
        \hline
        CON  & 5,163 & 4,420 & 3,623 & 2,578 & 4,109 & 5,260 & 5,297 \\ 
        \hline
        COFF & 2,792 & 3,656 & 4,469 & 5,271 & 7,072 & 6,127 & 7,184 \\ 
        \hline
        Total   & 7,955 & 8,076 & 8,092 & 7,849 & 11,181 & 11,387 & 12,481 \\ 
        \hline
        \hline
    \end{tabular}
    \label{tab:CON/COFF}
\end{table}

\subsubsection{The 2022 Reform: Introduction of Unified DAUC}
The reform removed the separated UC processes and replaced them with an integrated DAUC that simultaneously determines initial scheduling and price formation. The DAUC explicitly models the Total Transfer Capability (TTC) limits\footnote{TTC denotes the maximum reliable transfer capability over an interconnected transmission network subject to pre-and post-contingency constraints \citep{maliszewski1996available}.} between metropolitan areas (MA) and non-metropolitan areas (NMA) and enforces various system operational constraints directly into the optimization engine. 

From an operational perspective, this reform represented a significant advancement. Because the market-clearing schedules became physically feasible, the massive corrective actions previously required were minimized, which led to a substantial decline in out-of-market COFF payments \citep{__a1___2021}.

However, the SMP calculation methodology remains largely unchanged from the unconstrained merit-order process described in Section \ref{sec: price formation}. After the DAUC optimization determines the constrained dispatch schedules, a separate screening process identifies marginal units. Generators binding on transmission constraints (e.g., those constrained-off in NMA due to TTC limits) or stability constraints are excluded from setting the price. The SMP is then determined as the maximum stack price among the remaining marginal generators based on Eqs. (\ref{eq: SP}) and (\ref{eq: SMP}). This, however, constitutes a novel mismatch: despite incorporating transmission constraints into dispatch, the market price is still determined as a uniform SMP that obscures the locational value of these constraints. 

\begin{figure}
    \centering
	\includegraphics[scale = 0.28]{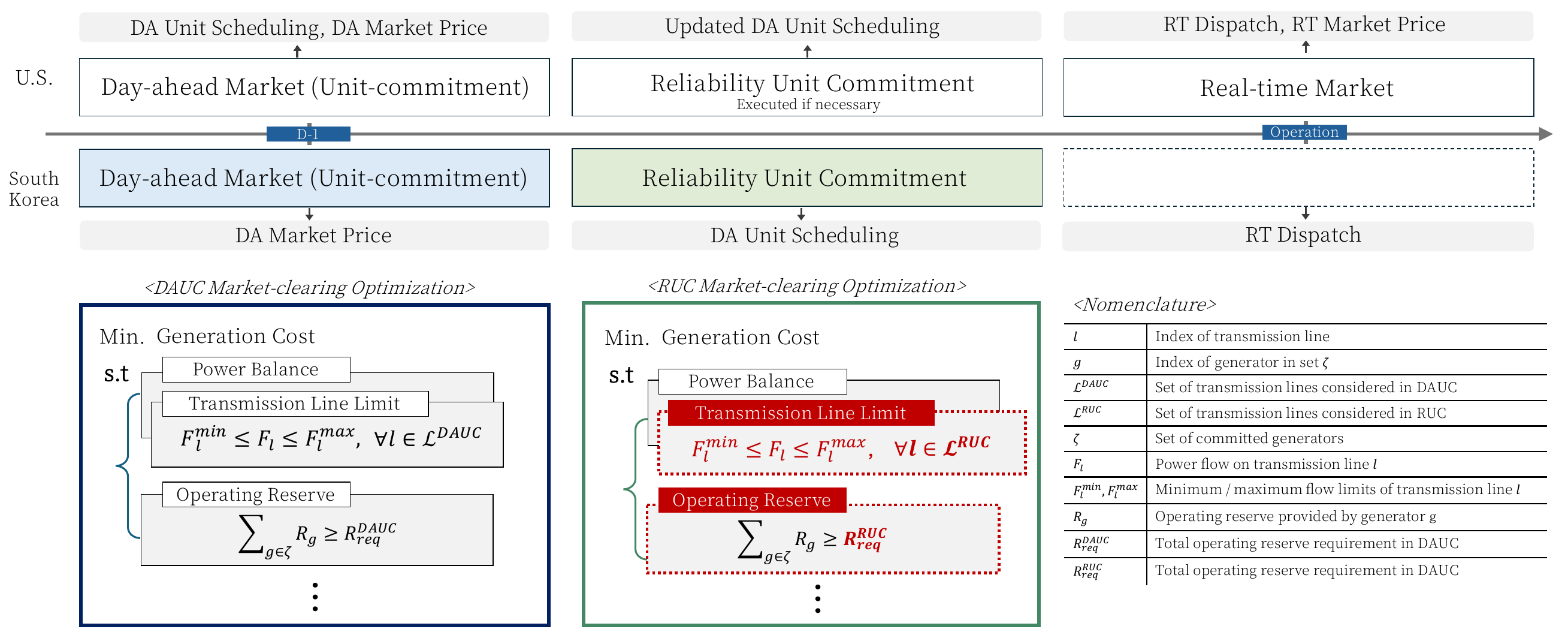}
	\caption{Misalignment between DAUC (Market Operation) and RUC (Power System Operation)}
	\label{fig:Discrepancy}
	\vspace{-1mm}
\end{figure}

\subsubsection{What the 2022 Reform Left Unresolved}
After DAUC closes, the system operator runs a Reliability Unit Commitment (RUC). The RUC establishes an updated scheduling reference for actual operation prior to real-time dispatch. While several U.S. ISOs also operate an RUC after the DAUC, it is not universally required; it is conducted only when system conditions necessitate it, such as forecast errors or unexpected outages\footnote{Their rules generally limit RUC to residual needs unmet by the day-ahead market, with any additional commitment solved on a least-cost basis.}. In practice, in markets such as PJM and ISO-NE, when a RUC action causes a resource to deviate from its DA scheduling in real-time, the resulting deviation is generally settled at the real-time price. By contrast, in the Korean market, the RUC operates differently. It is mandatory and runs unconditionally, redispatching all committed units under strict reliability criteria to ensure conservative system operation. 

Fig.~\ref{fig:Discrepancy} shows that the DAUC and RUC each consider different sets of constraints, including transmission limits, reserve requirements and minimum generation mandates for grid stability. The set of transmission lines considered in each process follows $\mathcal{L}^{DAUC} \subset \mathcal{L}^{RUC}$, which means that the RUC enforces transmission constraints on a broader set of lines. Reserve requirements also differ between the two processes; the RUC typically imposes a more conservative requirement, where $R_{req}^{RUC} > R_{req}^{DAUC}$. Consequently, this leads to systematic re-dispatch patterns, which are analyzed in detail in Section \ref{subsubsec:alignment}. 

The stack price-based screening approach from Section \ref{sec: price formation} was designed under a single-node assumption where transmission constraints are absent. While this remains applicable to the current two-node representation via TTC between MA and NMA, it cannot be extended to a meshed network topology. In meshed networks, locational value depends on how power at each location affects congestion across multiple interconnected lines. The merit-order screening approach becomes invalid under these conditions. Instead, prices must be determined using shadow prices of power balance and transmission constraints in the market clearing optimization.

\section{Diagnosing Price Formation Distortions in the Korean Electricity Market}\label{Sec3:Diagnosis}
The Korean power system is already experiencing observable operational stress despite the currently low penetration of renewable energy. These stresses arise from the spatial mismatch between demand-dominated areas and generation-dominated areas, the physical characteristics of an islanded grid, conservative operation to maintain stability, and a lack of flexible resources. However, the current market design does not provide price signals adequate to resolve these conditions or guide investment decisions.

Against this backdrop, the section first examines the physical background underlying operational stress in the Korean power system, then identifies how current price formation produces insufficient spatial and temporal signals, and finally describes the out-of-market interventions that have emerged as a result.

\subsection{Why the Korean Grid Struggles Despite Low Renewable Penetration} \label{subsec:Why}
Despite renewable energy accounting for 10\% of generation in 2023, which typically corresponds to IEA Phase 2 where variability is generally recognized as manageable \citep{solar2024wind, korea2025}, the Korean grid is already exhibiting operational problems characteristic of Phase 3 or 4, including frequent curtailment and localized stability issues. This premature onset of stress is driven by the following factors:

\subsubsection{Spatial Mismatch Between Supply and Demand-dominant Areas}
The MA (Seoul and Gyeonggi Province) accounts for roughly 41.1\% of national electricity demand but hosts only 30.3\% of total generation capacity \citep{KPX2024}. In contrast, most generation is concentrated in the southern coastal regions, with large nuclear and coal complexes along the East Coast and Southeast, while renewable energy, particularly solar PV, has expanded rapidly in Jeolla province (Southwest). This mismatch necessitates a northward flow of bulk power from generation-dominant regions to the load-intensive MA, placing heavy pressure on the transmission corridors connecting them.  

However, the usage of these transmission corridors is limited by N-1 reliability standards, which require the system to remain the loss of any single critical line. When power flows approach these limits, system operators must curtail generation output in supply-dominant regions to secure voltage and transient stability. Under such conditions, curtailment extends beyond VRE to inflexible baseload generators such as coal and sometimes nuclear units.

Expanding transmission capacity would be the most direct response to this bottleneck. In practice, however, transmission development in Korea has persistent delays due to social acceptance challenges, lengthy permitting processes, and regulatory barriers. For example, the 500 kV East Coast--MA transmission line, which was proposed to transfer power from the eastern coast to the capital region. The project encountered local opposition in 2009 and completed site selection only in early 2023. As a result, it is difficult to rely solely on transmission expansion as a timely solution to congestion.

\subsubsection{Isolated Grid Stability as a Constraint on Renewable Output}
Expanding the share of renewable generation in Korea faces limitations on two fronts, both of which require a higher online capacity of synchronous generators and thereby limit the scope for renewable dispatch.

First, unlike interconnected systems that can rely on neighbors for system stability support, Korea must provide all essential reliability services, such as inertia, frequency response, and frequency regulation, using domestic resources. This imposes a minimum online capacity of synchronous generators (coal and LNG) that must be maintained regardless of renewable availability or least-cost dispatch order, directly limiting the scope for renewable dispatch \citep{Im2024}.

Second, a substantial share of distributed photovoltaic resources currently installed in Korea were deployed under older grid codes that did not mandate Low Voltage Ride Through (LVRT) or Low Frequency Ride Through (LFRT) compliance. When a transmission-level fault occurs, this can  trigger the near-simultaneous involuntary disconnection of distributed PV assets across wide areas. Numerous studies have shown that such events can drive system frequency below critical security thresholds in the isolated grid \citep{9347727,Lee,JUNG20231374,SALEEM2024110184}. The system operator is accordingly obliged to maintain a minimum level of synchronous generation to secure sufficient spinning reserves against the risk of large-scale involuntary PV disconnection, leaving limited scope for expanding non-synchronous generation

\subsubsection{Insufficient Flexibility Resources} \label{subsubsec:flexibility resources}
Historically, the Korean power system prioritized economic efficiency and baseload stability. While this strategy succeeded in minimizing electricity costs, it resulted in a generation portfolio that was deficient in fast response resources. The Korean LNG fleet is dominated by large-scale combined-cycle units, and although these plants can operate in GT-only mode, such operation is not widely used as a primary source of operating flexibility\footnote{In 2025, documented fast-start deployments included GT-only starts within combined-cycle configurations to respond to sudden PV declines, but such use appears to remain limited.}.

The deployment of BESS, essential for shifting renewable energy and providing fast frequency response, has lagged behind the pace required for meaningful grid integration. Battery storage totals only 1.6 GW versus 15.8 GW (California) and 5.6 GW (Australia) \citep{park2023korean, KOTRA2024,CAISO2026}. Demand response capacity is similarly limited. Korea reports approximately 4.2 GW of obligated demand reduction capacity in its demand resource market \citep{KPX2025_DR}. However, these resources do not participate in the market to provide flexibility, and instead serve as emergency standby capacity held by the system operator. These inadequate flexibility resources fall far short of what is required to accommodate Korean renewable energy targets under the 11th Basic Plan, which calls for 77.2 GW of solar and 40.7 GW of wind by 2038.

\subsection{Lack of Locational Signal Provision} \label{subsec:spatial}
The Korean market accounts for network constraints in physical dispatch, but it still applies a uniform price across the mainland grid. The resulting divergence between physical operations and market clearing suppresses locational price signals, the key drivers for efficient resource allocation, and ultimately leads to misaligned incentives and operational inefficiencies. 

We investigate these signal deficiencies in South Korea in two ways. First, stylized network models are constructed to reflect this specific feature of the Korean market, where physical dispatch accounts for network constraints while market prices remain uniform. Second, simulations of the actual power grid are utilized to quantify the extent of locational signal distortions by comparing three distinct schemes\textemdash nodal, zonal, and uniform pricing.

\subsubsection{Theoretical Network Illustration} 
\begin{figure}
    \centering
	\includegraphics[scale = 0.4]{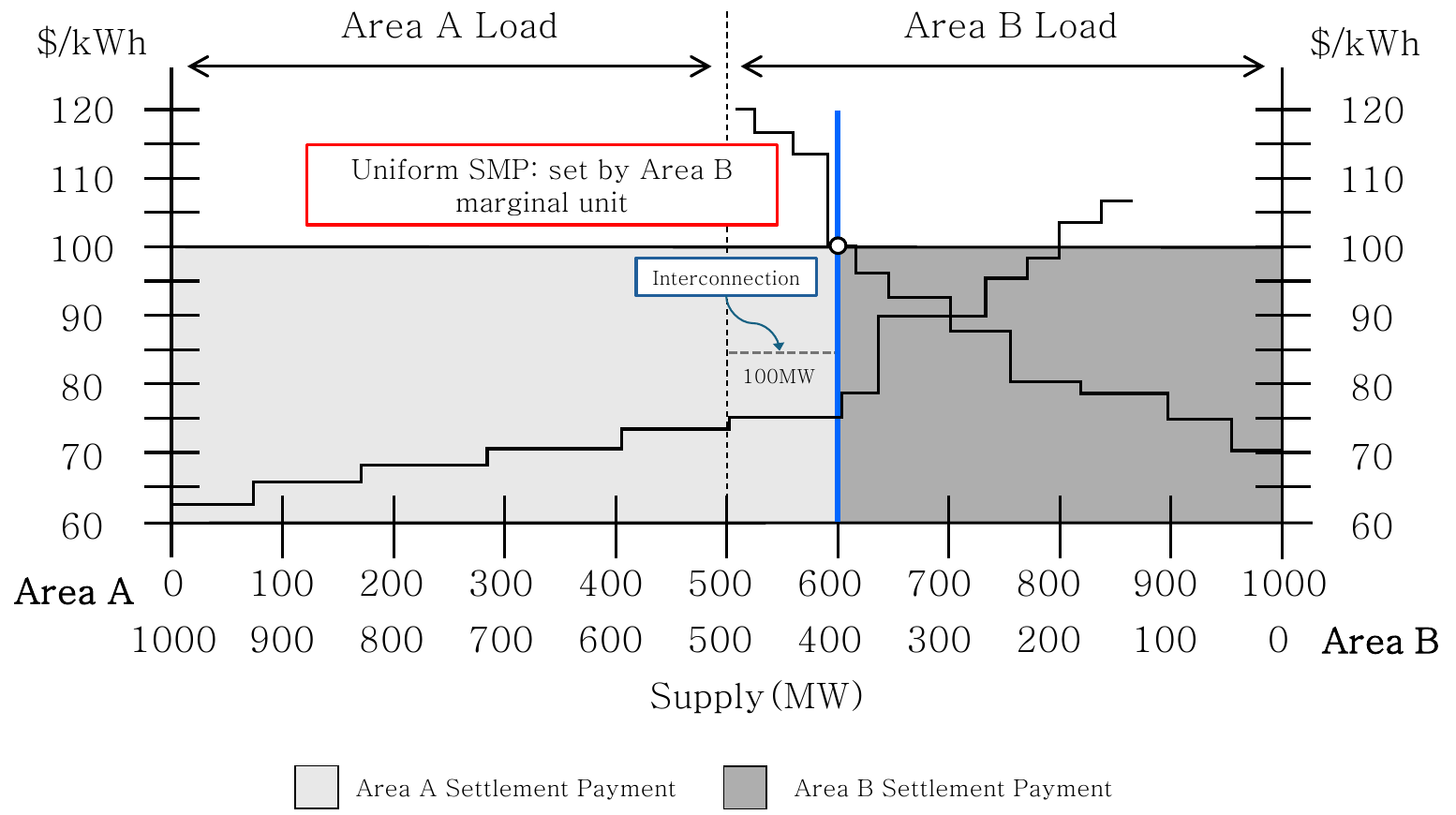}
	\caption{A Stylized Two-Bus Model Illustrating Price Distortion under Uniform Pricing}
	\label{fig:two-node example}
\end{figure}

\paragraph{(1) Inefficient Social Welfare Allocation}
Since the market-clearing engine enforces only the TTC between MA and NMA, abstracting complex local network constraints, the system can be effectively represented as a two-bus network for analytical purposes. Fig.~\ref{fig:two-node example} depicts this stylized model, focusing specifically on the interplay between transmission limits and price formation. The system parameters are defined as follows:

\begin{itemize}
    \item Area A (NMA): A generation-dominant region with a total capacity of 880 MW and a peak demand of 500 MW. The generation mix is dominated by nuclear and coal units, which result in relatively low marginal costs.
    \item Area B (MA): A load center with a total installed capacity of 420 MW and a peak demand of 500 MW. This zone depends on more expensive generation sources.
    \item Interconnection: The two zones are connected by a transmission interface with a capacity of 100 MW, representing the TTC constraint between MA and NMA.
\end{itemize}

In the standard uniform pricing arrangement found in the literature, market clearing assumes a single-node network that ignores transmission constraints, and physical dispatch is subsequently adjusted through congestion management. In Fig.~\ref{fig:two-node example}, if transmission constraints are absent, market clearing occurs where aggregate supply meets total demand of 1,000 MW, with Area A dispatching 700 MW and Area B 300 MW at a system-wide price of 90 KRW/kWh.

In contrast, the Korean market incorporates transmission constraints directly into the dispatch optimization, yet still applies a single uniform price. When the 100 MW transmission limit is reached, market price rises to 100 KRW/kWh, determined by the marginal unit in Area B. Although Area B generators are fairly compensated at this level, generators in Area A receive the same elevated price, well above their local marginal cost of 75 KRW/kWh. Since congestion costs are reflected in a single uniform price regardless of location, the true locational value of electricity is obscured, resulting in an inefficient distribution of social welfare.

\begin{figure}
    \centering
	\includegraphics[scale = 0.4]{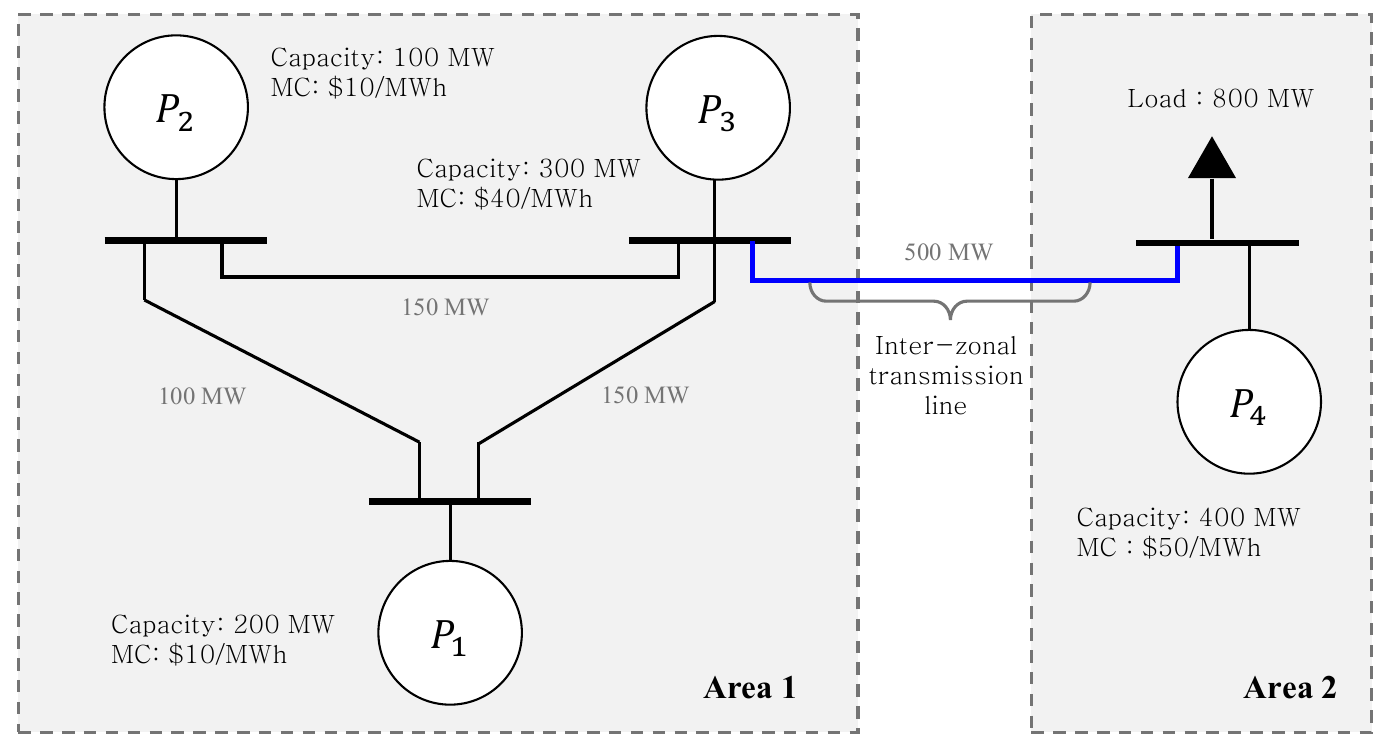}
	\caption{Four-node Toy Example}
	\label{fig:four-node example}
\end{figure}

\paragraph{(2) Dispatch Inefficiency under Inaccurate Network Representation} 
The accuracy with which transmission constraints are represented in market clearing directly determines how well prices reflect the physical value of the network. However, the Korean power system neither fully represents all transmission line constraints in its EMS nor directly incorporates them into market clearing. Instead, generator-side constraints are used in place of transmission constraints. We demonstrate how this leads to operational inefficiency using the four-node example in Fig.~\ref{fig:four-node example}.

\begin{itemize}
    \item Area 1: A generation-dominant zone with three generators ($P_1$, $P_2$, $P_3$) connected in a meshed network configuration, with capacities of 200, 100, and 800 MW and marginal costs of \$10, \$10, and \$40/MWh, respectively.
    \item Area 2: A load center with a single generator ($P_4$) of 400 MW capacity and a marginal cost of \$50/MWh, serving a demand of 800 MW, with the willingness to pay for demand set at \$100/MWh.
    \item Interconnection: The two areas are connected by a single inter-zonal transmission line with a capacity of 500 MW.
\end{itemize}

Table \ref{tab:placeholder_label} compares three cases: nodal representation incorporating all transmission constraints, zonal representation reflecting only the inter-zonal interconnection limit, and zonal representation with generator-side congestion management as practiced in the Korean market.

As shown in Table \ref{tab:placeholder_label}, nodal representation, derived from optimal power flow (OPF) incorporating all transmission constraints, yields the dispatch of (175, 100, 225, 300) MW that minimizes total generation cost. Although $P_1$ and $P_2$ share the same marginal cost, their nodal prices differ at \$10 and \$25/MWh, reflecting the locational marginal value of electricity under network constraints, with the highest price of \$50/MWh at Bus 4 where demand is concentrated.

When only the interconnection limit is reflected, intra-zonal line constraints are ignored and dispatch is determined in merit order within each zone, resulting in (200, 100, 200, 300) MW. Market prices are accordingly set at (\$40, \$50)/MWh. However, this dispatch is infeasible in actual system operation as line flows between buses 1--3 exceed their physical limits.

To prevent intra-zonal congestion, the Korean market instead uses generator-side constraints. When a minimum output constraint of 225 MW is imposed on $P_3$, the dispatch outcome matches that obtained under explicit transmission network representation. However, since $P_3$ is excluded from price formation due to the binding constraint, the Zone 1 price is set at \$10/MWh, the marginal cost of $P_1$, failing to capture the true locational value of electricity. This leads to an uplift payment of \$6,750, which raises total consumer payment from \$40,000 to \$46,750. Notably, the minimum output value of 225 MW in generator-side constraint assumed here represents an idealized optimum identical to the OPF solution. In practice, however, deriving the ideal constraint values across numerous generators under continuously changing system conditions is not feasible, and deviations from this optimum inevitably introduce further operational inefficiency.

\begin{table}[ht]
    \centering
    \caption{Impact of Transmission Network Representation on Market Clearing Outcomes}
    \label{tab:placeholder_label}
    \renewcommand{\arraystretch}{1.3}

    \resizebox{\textwidth}{!}{%
    \begin{tabular}{c|c|c|c}
        \hline
        \hline
        & \textbf{Nodal representation} & \textbf{Zonal} & \textbf{\makecell{Zonal \\ (Congestion Management)}} \\
        \hline
        \makecell{Generator\\Dispatch$^{a}$}
        & (175, 100, 225, 300)
        & (200, 100, 200, 300)
        & (175, 100, 225, 300) \\
        \hline
        Market Price$^{b}$
        & (10, 25, 40, 50)
        & (40, 50)
        & (10, 50) \\
        \hline
        Generator Revenue
        & \$28{,}250
        & \multirow{4}{*}{\raisebox{-0.8ex}{Not Available$^{c}$}}
        & \makecell{\$26{,}750 \\ (Uplift: \$6{,}750)} \\
        \cline{1-2}\cline{4-4}
        Consumer Payment
        & \$40{,}000
        &
        & \makecell{\$46{,}750 \\ (Uplift: \$6{,}750)} \\
        \cline{1-2}\cline{4-4}
        Congestion Rent
        & \$11{,}750
        &
        & \$20{,}000 \\
        \cline{1-2}\cline{4-4}
        Social Surplus$^{d}$
        & \$53{,}250
        &
        & \$53{,}250 \\
        \hline
        \hline
    \end{tabular}%
    }

    \vspace{0.3em}
    \begin{minipage}{\textwidth}
    \footnotesize
    $^{a}$Generator dispatch is listed in the order of ($P_1$, $P_2$, $P_3$, $P_4$).\\
    $^{b}$Market prices are determined at the bus level under nodal representation and at the zonal level under zonal pricing.\\
    $^{c}$Not available because intra-zonal line flows exceed physical limits, causing this solution to be infeasible in actual system operation.\\
    $^{d}$Social surplus is defined as the sum of producer surplus, consumer surplus, and congestion rent, where producer surplus is the difference between market revenue and production cost, and consumer surplus is the difference between total consumer utility and electricity payment.
    \end{minipage}
\end{table}

\begin{table}[t]
    \centering
    \caption{Uniform and Zonal Price Result (KRW/kWh)}
    \renewcommand{\arraystretch}{1.3}
    \begin{tabular}{c||c|c|c|c|c|c}
        \hline
        \hline
        \multicolumn{3}{c|}{Category} & Morning & Daytime & Evening & Night \\
        \hline
        \multirow{3}{*}{Spring} 
         & \multicolumn{2}{c|}{Uniform} & 123.10 & 124.76 & 140.26 & 123.10 \\
        \cline{2-7}
         & \multirow{2}{*}{Zonal}
         & MA & 123.10 & 124.76 & 140.26 & 123.10 \\
        \cline{3-7}
         &    & NMA    & 123.10 & 114.71 & 140.26 & 123.10 \\
        \hline
        \multirow{3}{*}{Summer} 
         & \multicolumn{2}{c|}{Uniform} & 123.37 & 142.48 & 143.44 & 124.71 \\
        \cline{2-7}
         & \multirow{2}{*}{Zonal}
         & MA & 123.37 & 142.48 & 143.44 & 124.71 \\
        \cline{3-7}
         &    & NMA    & 100.05 & 130.28 & 142.49 & 100.23 \\
        \hline
        \multirow{3}{*}{Fall} 
         & \multicolumn{2}{c|}{Uniform} & 124.81 & 133.14 & 143.12 & 124.51 \\
        \cline{2-7}
         & \multirow{2}{*}{Zonal}
         & MA & 124.81 & 133.14 & 143.12 & 124.51 \\
        \cline{3-7}
         &    & NMA    & 113.42 & 116.81 & 143.12 & 111.22 \\
        \hline
        \multirow{3}{*}{Winter} 
         & \multicolumn{2}{c|}{Uniform} & 139.62 & 142.48 & 144.04 & 132.22 \\
        \cline{2-7}
         & \multirow{2}{*}{Zonal}
         & MA & 139.62 & 142.48 & 144.04 & 132.22 \\
        \cline{3-7}
         &    & NMA    & 118.33 & 131.89 & 141.95 & 114.50 \\
    \hline
    \hline
    \end{tabular}
    \label{tab:Uniform and Zonal Price Result (KRW/kWh)}
\end{table}

\begin{table}
    \centering
    \caption{Case Study Simulation Details}
    \renewcommand{\arraystretch}{1.3}
    \resizebox{\textwidth}{!}{%
    \begin{tabular}{c|c}
        \hline
        \hline
         \textbf{Category} & \textbf{Description} \\
         \hline
         Dispatch Method & Security-Constrained Economic Dispatch (SCED)\\
         \hline
         Demand & 2023 Actual Profile\\
         \hline
         Load Zones & \makecell{ Zonal: MA/NMA; \\ Nodal: nodal demand derived form 10th Basic Plan)} \\
         \hline
         Generator Capacity & 10th Basic Plan \\
         \hline
         Renewable Output & Historical 2023 Profiles \\
         \hline
         Scenarios & \makecell{16 scenarios: 4 seasons (spring, summer, fall, winter) \\ × 4 time-of-day (night, morning, daytime, evening)}\\
         \hline
         \hline
    \end{tabular}}
    \label{tab:placeholder}
\end{table}

\subsubsection{Case Study: Actual Bulk Power Systems} \label{subsubsec:bulk power}
To examine how transmission constraints shape locational prices in the Korean system, we model the bulk power system in PLEXOS 10.0 using a 4,240-bus network representation of the actual Korean power system. For each scenario, a security-constrained economic dispatch is performed to minimize total generation cost subject to transmission flow limits, generator private output constraints, and system constraints. The market price is derived from the dual variable of the power balance and transmission flow constraints, and does not account for quasi-fixed costs such as start-up and no-load costs. Simulation details are summarized in Table \ref{tab:placeholder}, and further details regarding the model and input data are described in \ref{app2}.

The Korean power system exhibits meaningful spatial price differences that are entirely suppressed under the current uniform pricing mechanism. The simulation results quantify both the extent of this suppression and its consequences for investment signals.

Under uniform pricing, the national SMP ranges from 123.10 to 144.04 KRW/kWh across the 16 scenarios, varying by season and time of day but producing no spatial differentiation. As in the example illustrated in Fig.\ref{fig:two-node example}, NMA generators are overcompensated despite having their output restricted by transmission bottlenecks, while NMA consumers pay rates that significantly exceed their true locational marginal cost.

When prices are separated by zone, MA prices are on average 10.16 KRW/kWh higher than NMA prices, indicating that transmission constraints give rise to spatial price gaps. This gap widens to 14.1 KRW/kWh during high-demand summer and winter periods, and narrows to 6.3 KRW/kWh during low-demand spring and fall scenarios when constraints rarely bind.

\begin{figure}
    \centering
	\includegraphics[scale = 0.12]{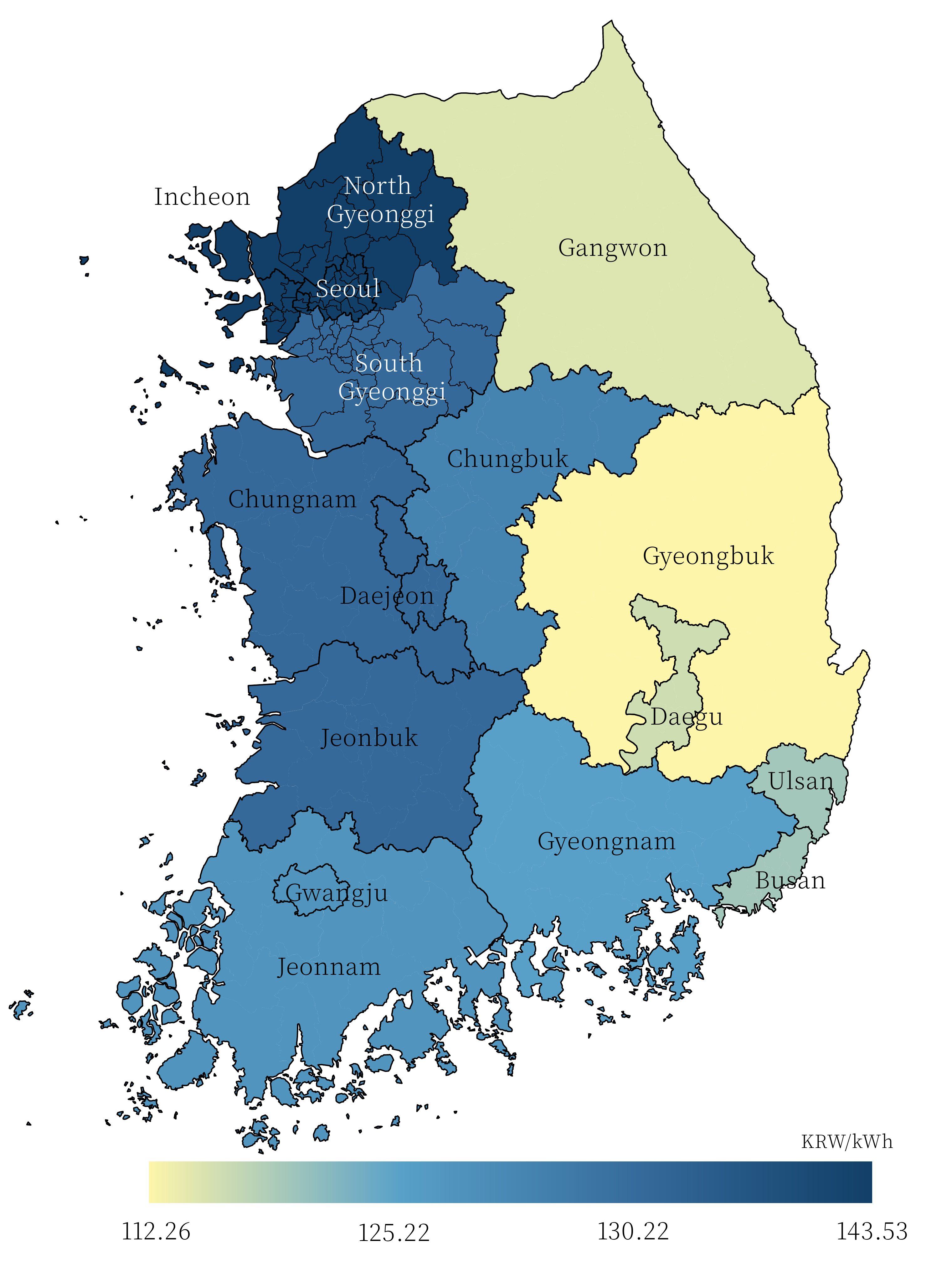}
	\caption{LMP Simulation Result on Korean Power System}
	\label{fig:LMP result}
\end{figure}

\begin{table}[t]
    \centering
    \caption{Nodal Price Result of (Fall, Daytime) (KRW/kWh)}
    \renewcommand{\arraystretch}{1.4}
    \begin{tabular}{c|c||c|c|c}
        \hline
        \hline
        \diagbox[width=6em]{\textbf{Province}}{\textbf{Price}} & LMP & \makecell{Energy \\ Component} & \makecell{Congestion\\Component} & \makecell{Loss \\ Component} \\
        \hline
        \makecell{Zone\,1 \\[-0.4ex] {\scriptsize (Metropolitan)}} & 142.1 & \multirow{5}{*}{\raisebox{-0.8ex}{124.04}} & 16.42 & 1.64 \\
        \cline{1-2}
        \cline{4-5}
        \makecell{Zone\,2 \\[-0.4ex] {\scriptsize (Chungcheong)}} & 131.9 & & 6.51 & 1.44 \\
        \cline{1-2}
        \cline{4-5}
        \makecell{Zone\,3 \\[-0.4ex] {\scriptsize (Jeolla)}} & 127.7 & & 3.97 & -0.25 \\
        \cline{1-2}
        \cline{4-5}
        \makecell{Zone\,4 \\[-0.4ex] {\scriptsize (Gyeongsang)}} & 121.7 & & -3.06 & 0.77 \\
        \cline{1-2}
        \cline{4-5}
        \makecell{Zone\,5 \\[-0.4ex] {\scriptsize (Gangwon)}} & 114.3 & & -8.48 & -1.17 \\
        \hline
        \hline
    \end{tabular}
    \label{tab:Nodal Price Result}
\end{table}

At the nodal level, this divergence becomes even more pronounced. Fig.~\ref{fig:LMP result} displays the geographic distribution of nodal prices for the representative fall daytime scenario, with the highest prices concentrated in the MA and the lowest in Zone 4 (Gyeongsang Province) and Zone 5 (Gangwon Province).

Table \ref{tab:Nodal Price Result} decomposes nodal prices into energy, loss, and congestion components. Although prices are calculated at the nodal level, the results are aggregated into five administrative regions to facilitate interpretation. The congestion component ranges from +16.42 KRW/kWh in Zone 1 (MA) to -8.48 KRW/kWh in Zone 5 (Gangwon Province), a spread of 24.9 KRW/kWh. The positive congestion component in Zone 1 indicates that meeting incremental demand in the MA necessitates dispatching high-cost generators located within the MA due to transmission import limits. Conversely, the negative congestion component in Zone 5 signifies that generation in Gangwon province exacerbates transmission bottlenecks. This reduces the marginal value of electricity at these nodes by 8.48 KRW/kWh below the system average. As a result, total nodal prices span 27.8 KRW/kWh across zones\textemdash a gap entirely obscured under uniform pricing.

Under locational pricing, price gaps of this magnitude would provide direct economic incentives for market participants to site generation where it is most needed: higher prices in Zone 1 would attract investment toward major load centers, while lower prices in transmission-constrained regions such as Zone 3 would discourage further concentration of renewable capacity. Instead, the uniform price masks these signals, perpetuating the spatial mismatch between generation and load.

\subsection{Temporal Signal Deficiency: Absence of Real-Time Market and CBP} \label{subsec:temporal}
In standard two-settlement systems \citep{HOHL2023113503}, a real-time market with sub-hourly granularity (5 to 15 minutes) corrects day-ahead schedules and compensates fast response resources. In Korea, however, day-ahead prices are applied directly to real-time output ($SMP_{DA} \times Q_{RT}$) on an hourly basis, and deviations are handled through administrative mechanisms such as MWP and DAMAP outside the market clearing process. Furthermore, since generators do not submit price offers under CBP, the SMP is determined solely based on the regulated variable costs of the marginal unit, failing to capture opportunity costs and the value of scarcity.

\begin{figure}[htbp]
    \centering
    \begin{subfigure}[b]{0.7\textwidth}
        \centering
        \includegraphics[width=\linewidth]{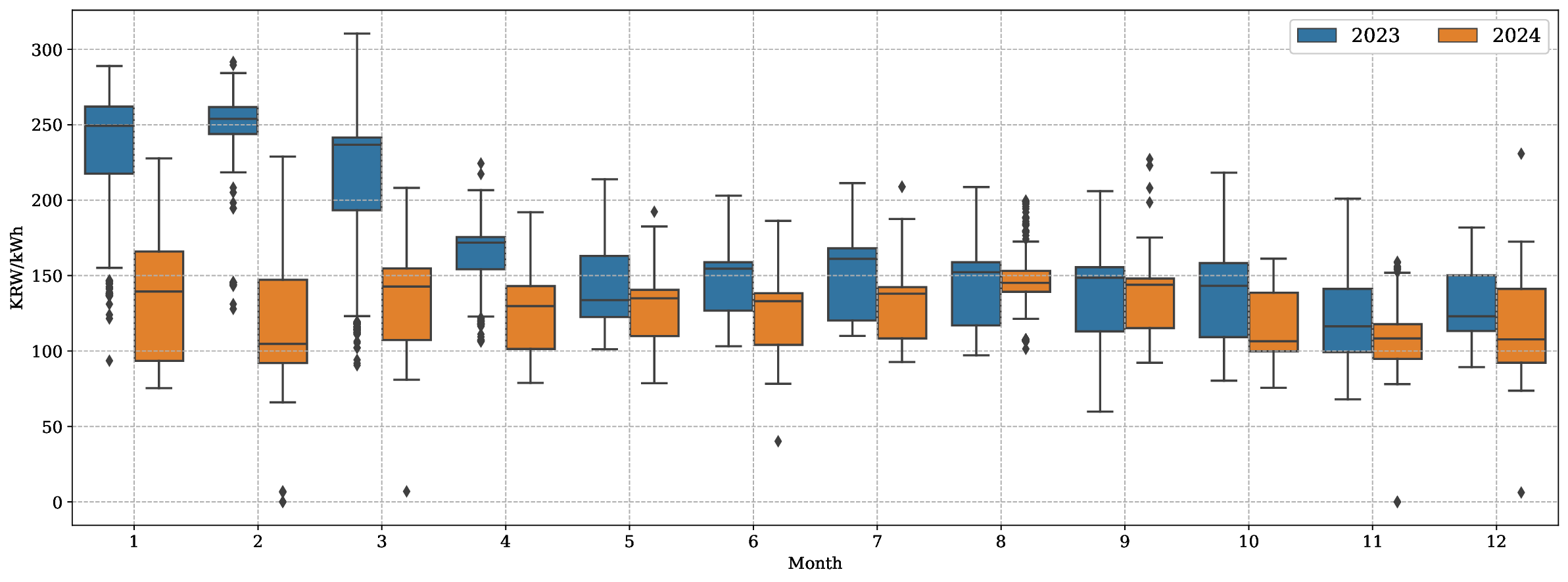}
        \caption{Monthly Boxplot of SMP for 2023 and 2024}
        \label{fig:SMPBoxplot}
    \end{subfigure}
    \vspace{1em} 
    \begin{subfigure}[b]{0.7\textwidth}
        \centering
        \includegraphics[width=\linewidth]{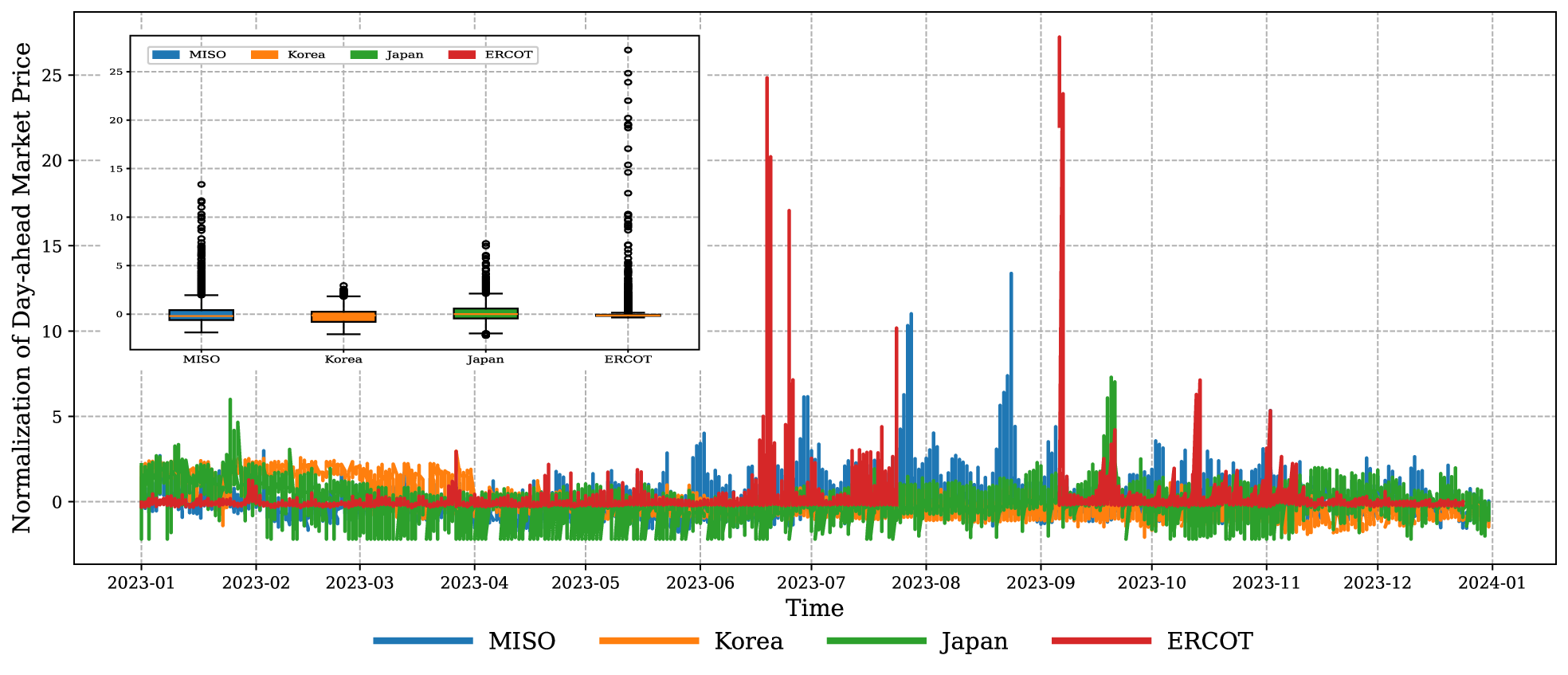}
        \caption{Normalized day-ahead Market Price in 2023 (ERCOT, MISO, Japan, Korea)}
        \label{fig:NormalizedDAMPrice}
    \end{subfigure}

    \caption{Comparative Analysis of Electricity Market Price Patterns}
    \label{fig:comparison-vertical}
\end{figure}

The consequences are evident in observed price patterns. As shown in Fig.~\ref{fig:comparison-vertical}, Korean SMP remained within a narrow range over 2023--2024, with a median around 133.12 KRW/kWh and a 90th percentile of approximately 155.48 KRW/kWh. This limited variation contrasts sharply with ERCOT, which exhibits frequent scarcity-driven price spikes, and MISO and JEPX, which demonstrate significantly higher price volatility. Such low price volatility, caused by the absence of competitive bidding and real-time pricing, cannot provide sufficient incentives for the flexible resources that the system needs, and the resulting shortage of these resources in Korea has already been documented in Section \ref{subsubsec:flexibility resources}.

\subsection{Out-of-Market Actions in Response to Pricing Failures} \label{Subsec:Administrative}
Market prices in Korea do not reflect the true spatial and temporal value of electricity, leaving the operational challenges identified in Section \ref{subsec:Why} unresolved. As a result, the Korean market has increasingly relied on government-dominated interventions aimed at addressing immediate symptoms rather than underlying causes. Three specific measures are analyzed in this section: the suspension of new grid interconnections, financial compensation for renewable energy curtailment, and the centralized procurement of flexibility resources.

\subsubsection{Suspension of New Interconnections in Jeolla Province} \label{subsubsec:Jeolla}
\begin{figure}[htbp]
    \centering
    \begin{subfigure}[b]{0.4\textwidth}
        \centering
        \includegraphics[width=\linewidth]{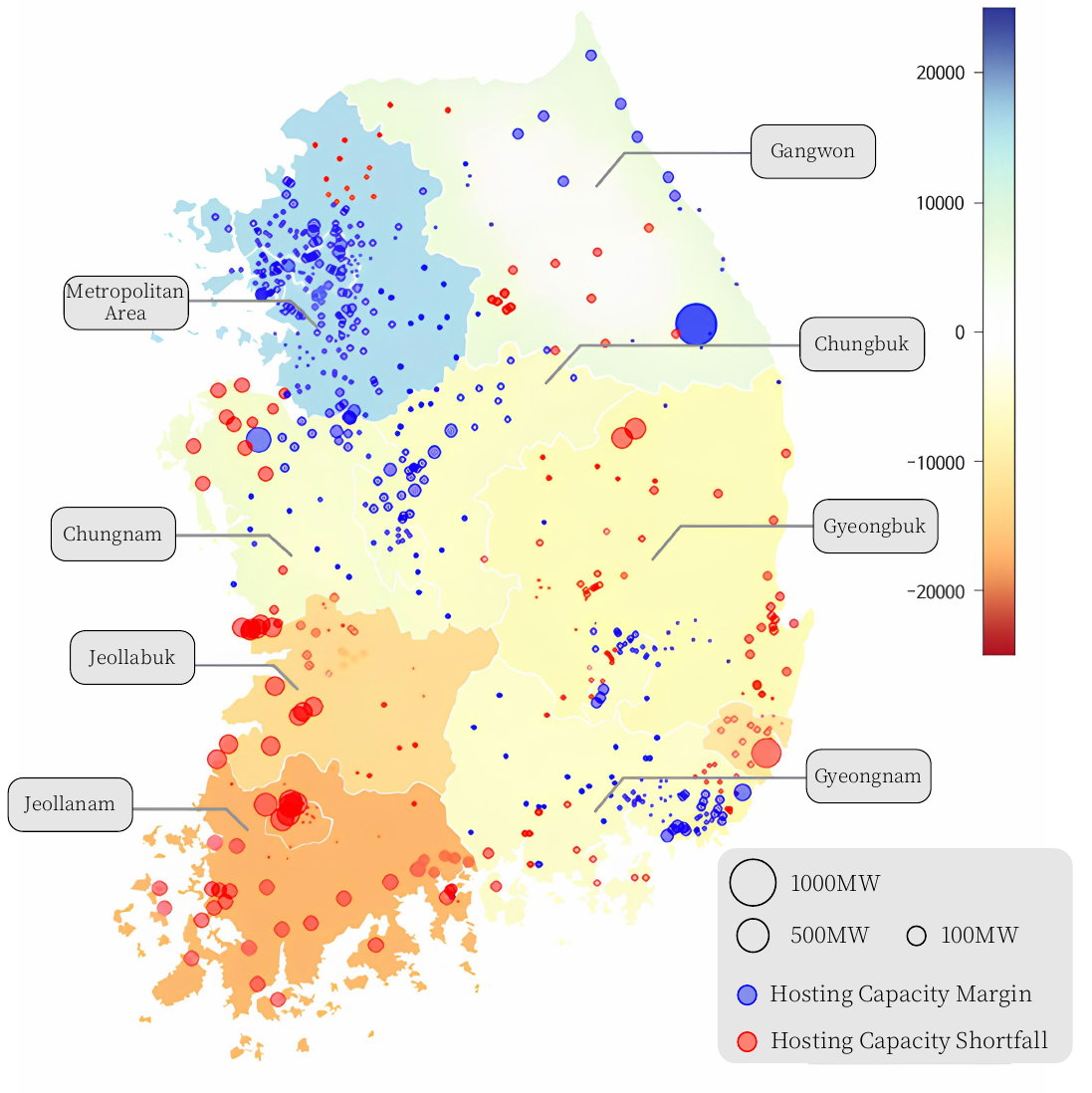}
        \caption{2026-2030}
        \label{fig:HostingCap_2630}
    \end{subfigure}
    \begin{subfigure}[b]{0.4\textwidth}
        \centering
        \includegraphics[width=\linewidth]{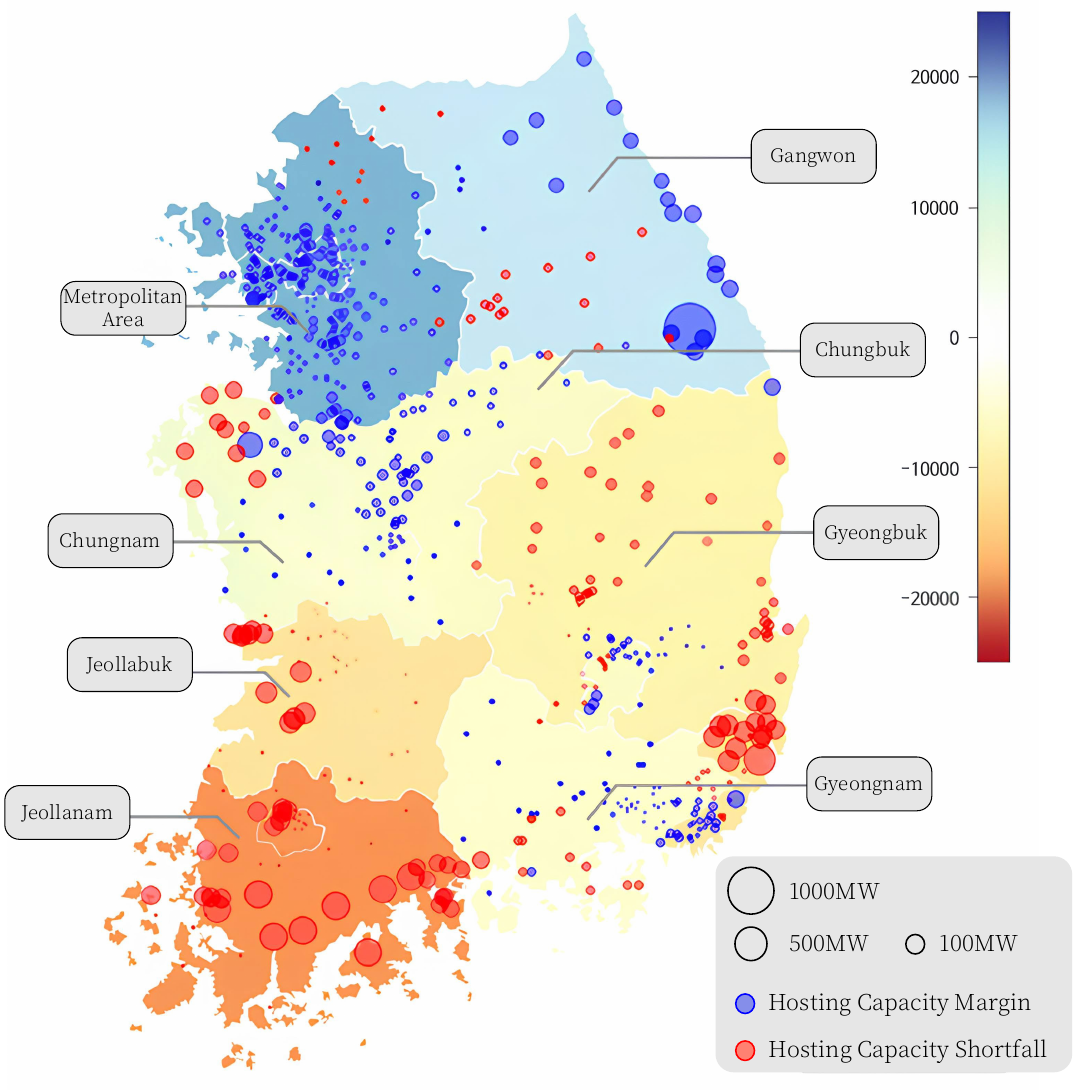}
        \caption{2031-2035}
        \label{fig:HostingCap_3135}
    \end{subfigure}

    \caption{Spatial distribution of hosting capacity margins and shortfalls across the Korean power system}
    \label{fig:horizontal}
\end{figure}

Jeolla province has become a major hub for PV investment, largely due to its abundant solar irradiance and relatively low land costs. It accounts for roughly 32.28\% of national renewable capacity. The 2024 hosting capacity analysis by KPX indicates that substations across Jeolla province have exhausted available connection margins. As illustrated in Fig.~\ref{fig:horizontal}, hosting capacity shortfalls are heavily concentrated in Jeolla province across both projection periods (2026--2030 and 2031--2035). Substations across Jeollanam show capacity shortfalls in all monitored locations, while North Jeollabuk similarly reports shortfalls at the majority of substations in both periods. Consequently, the government suspended new renewable interconnections in the region until approximately 2032. The suspension reduces the immediate risk of transmission congestion, but significantly constrains future renewable expansion in the region that hosts the largest share of national renewable capacity. 

\subsubsection{Administrative Payments for Curtailment}
As concerns over renewable curtailment have grown, the government introduced a compensation mechanism known domestically as quasi-central dispatch. Renewable generators participating in this scheme submit day-ahead bids and are subject to dispatch instructions from the system operator. When instructed to curtail, they are compensated approximately 10.6 KRW/kWh for their scheduled generation volumes, intended to cover the opportunity costs incurred by complying with dispatch instructions. Notably, the scheme currently operates only during spring, reflecting the seasonal concentration of curtailment driven by excess PV during low-demand periods. Furthermore, participation is restricted to generators connected to substations in the Honam region, consistent with the hosting capacity shortfalls identified in Section \ref{subsubsec:Jeolla}. 

However, this administrative compensation may inadvertently reinforce the very problem it seeks to address. By guaranteeing payment for curtailed generation, it reduces the financial risk of siting in transmission-constrained areas, potentially encouraging further concentration of renewable investment in already congested regions. 

\subsubsection{Centralized BESS Auctions}
The absence of a real-time market and scarcity price signals, as discussed in Section \ref{subsec:temporal}, means that the spot market provides insufficient incentives for storage investment. In response, the Korean government launched a centralized capacity auction, a tender-based procurement approach, to secure a predetermined volume of flexibility resources. In the first auction round in 2025, a total of 540 MW of capacity was procured (500 MW on the mainland and 40 MW on Jeju) for the 2026 requirement, followed by a second auction for the 2027 requirement.

However, this procurement design faces several concerns regarding economic efficiency and transparency. While an ideal market should remain technology-neutral to allow diverse flexibility providers, such as demand response or alternative long-duration storage, to compete, the current tender is strictly technology-specific, limited only to BESS. This exclusionary approach prevents the discovery of the most cost-effective flexibility mix. The basis for the specific procurement volumes is also not publicly disclosed: although these figures are derived from the 11th Basic Plan for Electricity Supply and Demand, the analytical basis for translating planning targets into specific auction volumes has not been established transparently, raising concerns about potential over-investment or misalignment with actual grid requirements.

Beyond the procurement design, significant concerns remain regarding operational efficiency. The operational value of storage varies considerably across time depending on system conditions, and capturing this value requires price-responsive scheduling. Under the current mainland contract, however, BESS operation is governed by system operator dispatch instructions rather than self-scheduling based on market price signals. Settlement is calculated using available capacity and implementation rates, meaning contracted resources have no direct financial incentive to optimize their charging and discharging cycles.

The auction design further limits the broader utility of these assets by prioritizing grid-congested areas. By assigning higher scores to bidders siting assets near curtailment-heavy substations, the design concentrates resources at specific nodes, potentially undermining their ability to provide wider system services such as frequency regulation or peak shaving across the entire grid.

\begin{landscape}
\begin{table}[p]
    \centering
    \caption{Why the institutional configuration of the Korean electricity market gives rise to distorted price formation}
    \label{tab:summary}
    \renewcommand{\arraystretch}{1.25}
    \small
    \hyphenpenalty=10000
    \exhyphenpenalty=10000

    \begin{tabular}{
        >{\RaggedRight\arraybackslash}p{2.8cm}|
        >{\RaggedRight\arraybackslash}p{5.6cm}|
        >{\RaggedRight\arraybackslash}p{5.6cm}|
        >{\RaggedRight\arraybackslash}p{5.6cm}
    }
        \hline
        \hline
        \textbf{Generic market design issue}
        & \textbf{How this issue is commonly addressed in mature markets}
        & \textbf{Korea-specific institutional feature}
        & \textbf{Result of market failure in Korea} \\
        \hline

        Pricing of transmission congestion
        & Congestion is typically reflected through nodal or zonal pricing
        & Physical dispatch reflects network constraints, but price is based on a nationwide uniform SMP
        & The marginal value of congestion is suppressed in prices, distorting locational signals and siting incentives \\
        \hline

        Reflection of a real-time system conditions in prices
        & Mature markets typically use real-time markets or balancing mechanisms to reflect short-run scarcity and real-time conditions
        & Korea operates only a day-ahead market
        & Operational and investment signals for flexible resources are weakened by prices that fail to capture real-time system conditions \\
        \hline

        Offer-based price formation
        & Market participants submit quantity-price offers, and prices are set based on these offers through market clearing
        & Under the CBP, Korea relies on administratively assessed costs rather than participant-submitted price-quantity offers
        & By depending entirely on administratively assessed costs, the market suppresses price volatility and weakens the signals needed to guide operational and investment decisions \\
        \hline

        Consistency between market prices and dispatch quantities
        & Standard market design emphasizes market prices compatible with dispatch
        & Market prices are determined day-ahead, while settlement follows actual dispatch during real-time operation
        & The separation between system operation and price formation gives rise to out-of-market uplift payments from redispatch gaps \\
        \hline
        \hline
    \end{tabular}
\end{table}
\end{landscape}

\section{Reform Tasks for Restoring the Price Function} \label{Sec4:Package}
This section proposes a reform package aimed at replacing out-of-market administrative interventions with effective price signals, addressing the market design deficiencies identified in the Korean electricity market and summarized in Table \ref{tab:summary}. It is organized in two parts: key market design reforms are identified first, followed by a sequenced implementation pathway. 

\subsection{Key Market Reform Proposals}
\subsubsection{Adoption of Locational Marginal Pricing} \label{subsubsec:LMP}

\begin{figure}
    \centering
	\includegraphics[scale = 0.4]{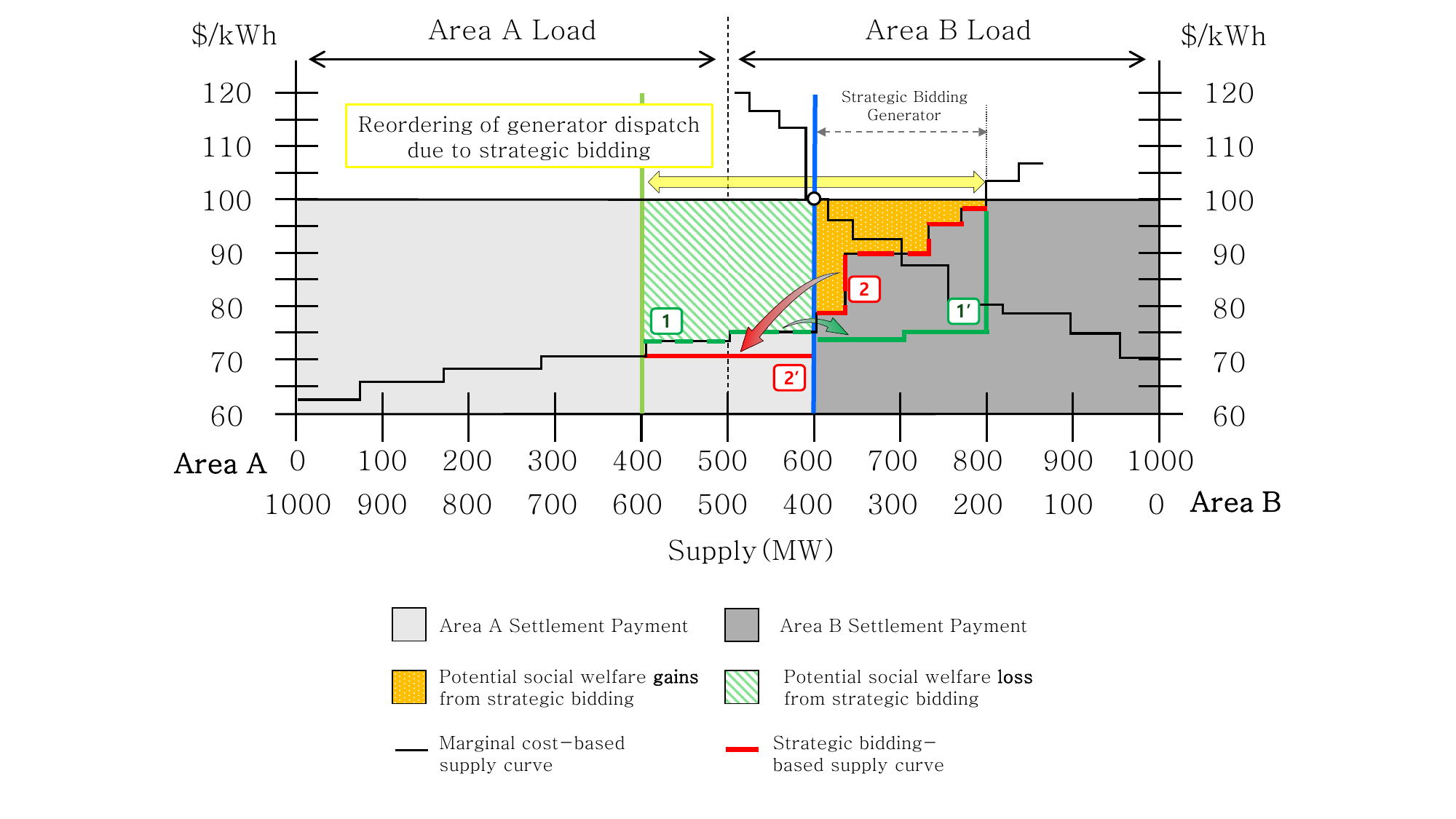}
	\vspace{-3mm}
	\caption{Strategic Bidding with Inaccurate Transmission Representation}
	\label{fig:gaming in KR}
	\vspace{-1mm}
\end{figure}

The inefficient allocation of social welfare resulting from inaccurate price signals under uniform pricing, as described in Section \ref{subsec:spatial}, provides generators in certain locations with opportunities for strategic bidding, once generators are allowed to submit individual price offers.

Fig.~\ref{fig:gaming in KR} illustrates this mechanisms. Under the current arrangement, the transmission constraint between Area A and Area B means that the market-clearing price is always set by high-cost units in Area B, regardless of how Area A generators bid. Generators in Area A with marginal costs between 75 and 100 KRW/kWh are not cleared under truthful bidding, as illustrated by the black supply curve. However, by submitting bids below their true marginal cost, as shown by the red supply curve, these generators can enter the dispatch order and be settled at the market price of 100 KRW/kWh, capturing the additional surplus shown in yellow. This reordering displaces generators that bid truthfully, resulting in the social welfare loss indicated in green. Strategic bidding of this kind reduces market efficiency when bids do not reflect least-cost production \citep{hortaccsu2008understanding}.

The current SMP yields inaccurate price signals and, as demonstrated above, provides opportunities for strategic bidding. This arises from the inability of uniform pricing to reflect the scarcity value of transmission capacity. Introducing locational pricing would address this by incorporating the shadow price of transmission constraints to market prices.

Korea is currently considering a gradual transition toward locational pricing, beginning with zonal pricing as an initial step before moving to pricing at the nodal level. When introducing zonal pricing, the primary consideration is how to draw the zone boundaries. In the Korean context, zone boundaries should be based on the electrical characteristics. In markets such as those in Europe and Japan, zone boundaries have historically followed administrative divisions, reflecting the vertically integrated regional frameworks from which those markets evolved. Korea, by contrast, has developed its power industry under direct state oversight, with both system and market operations conducted through centralized optimization.

The current system operation practice provides a reference for determining zone boundaries based on electrical characteristics. The existing DAUC already incorporates the interconnection limit between the MA and NMA as the only transmission constraint in determining generator dispatch, indicating that this interface functions as the effective congestion boundary in the Korean system. Given that locational pricing is intended to reflect the value of transmission congestion, adopting this boundary ensures consistency between dispatch outcomes and price formation\textemdash aligned with the intent of the DAUC.

\subsubsection{Introduction of Real-Time Market}
Real-time system operation inevitably involves imbalances arising from generator outages, ramp rate limitations, renewable intermittency, demand forecast errors, and transmission contingencies. As a result, the system operator must continuously adjust generation output in real time and, when necessary, secure additional resources at short notice. Without a market mechanism that reflects these conditions in prices, the time-varying marginal value of electricity does not captured, leading to the limited price volatility identified in Section \ref{subsec:temporal}. This, in turn, removes the market incentives needed for flexible resources to participate\textemdash a shortage already documented in Section \ref{subsubsec:flexibility resources}.

A real-time market addresses this by generating price signals that align with actual dispatch outcomes, reflecting the marginal value of electricity as it varies with real-time supply and demand conditions. Beyond its immediate role in system balancing, efficient real-time pricing also impacts on incentives in day-ahead and forward markets \citep{HOGAN201423}. Day-ahead prices are shaped by expectations of real-time prices, and persistent deviations between day-ahead and expected real-time prices create arbitrage opportunities \citep{HOGAN201633}. Real-time prices also serve as a reference for forward contract pricing, since forward prices reflect expectations of future spot-market conditions \citep{HOGAN201423}.

In established power markets, the necessity of real-time markets has long been recognized. FERC Order No. 2000, issued in 1999, mandated real-time balancing markets across all U.S. RTOs, affirming that such markets are essential for nondiscriminatory grid access and competitive energy market development \citep{hirst2001real}. Real-time markets are currently operated across all U.S. ISOs/RTOs and other major markets worldwide, including Canada, New Zealand, and Australia.

\begin{figure}
    \centering
	\includegraphics[scale = 0.35]{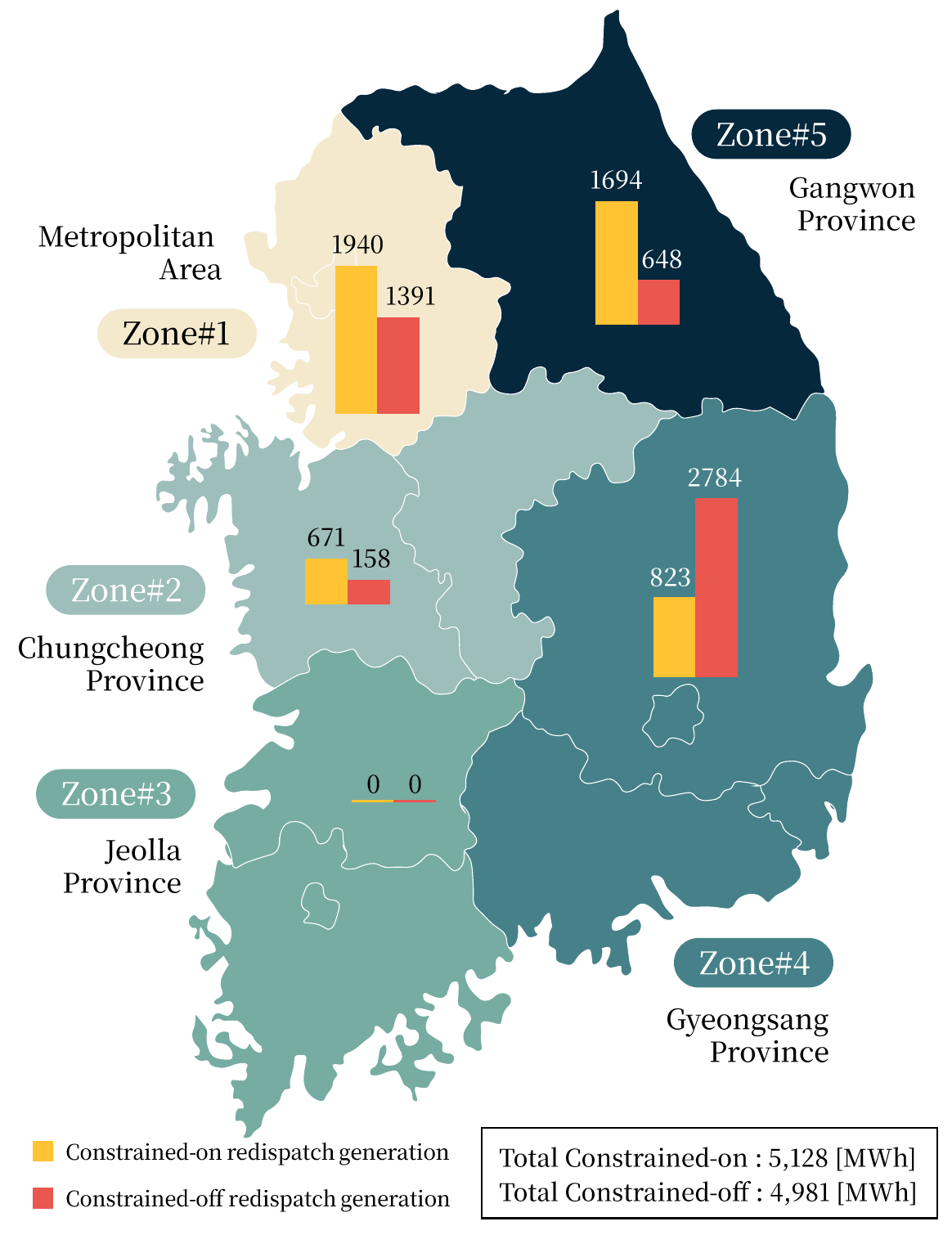}
	\vspace{-3mm}
	\caption{Geographical Distribution of Generation Redispatch}
	\label{fig:Result1}
	\vspace{-1mm}
\end{figure}

\subsubsection{Aligning Dispatch and Pricing through EMS Enhancement} \label{subsubsec:alignment}
In standard ISO arrangements, market operations are integrated, with SCED simultaneously determining prices and dispatch in real time. In Korea, however, the market is operated exclusively through DAUC, and the centralized optimization prior to real-time dispatch is carried out through RUC. Notably, as described above, RUC is designed for actual system operation and therefore operates under a different optimization model from DAUC. In other words, market operations and system operations are decoupled in Korea.

The difference between the generation schedule determined in DAUC and the actual dispatch carried out through RUC is referred to as redispatch. Given the structural differences between the two models discussed above, this redispatch is expected to exhibit consistent and asymmetric patterns across regions rather than random deviations. To empirically verify this, we simulated the sequential DAUC and RUC processes using the bulk power system model described in Section \ref{subsubsec:bulk power}.

As shown in Fig.~\ref{fig:Result1} for the (Fall,Daytime) scenario, Zone 1 is dominated by constrained-on generation while Zone 4 exhibits constrained-off generation. Table \ref{tab:Redispatch result1} further shows that this pattern holds consistently across both (Fall, Daytime) and (Spring, Daytime) scenarios, selected as representative scenarios of low-demand, high-renewable conditions.

In the DAUC, where operational constraints are less stringent as illustrated in Fig.~\ref{fig:Discrepancy}, lower-cost generators in Zone 4, predominantly nuclear and coal units, are dispatched at higher output levels. However, as the RUC incorporates additional operational constraints to ensure system stability and prevent line overloads, generation output of Zone 4 is reduced. The resulting gap is covered by higher-cost LNG generators in Zone 1.

To bridge this gap, aligning the EMS models underlying market clearing and generation scheduling is essential. Without this alignment, price signals produced by any reformed pricing mechanism will continue to diverge from actual system operation, limiting the effectiveness of other reform packages discussed in Section \ref{Sec4:Package}. In particular, these predictable redispatch patterns could incentivize strategic bidding if a price-based bidding system were introduced under the current EMS structure. Generators in persistently COFF regions such as Zone 4 can anticipate that their output will be reduced in RUC regardless of their day-ahead schedules. This creates an incentive to bid low in the day-ahead market, as a larger scheduled quantity results in greater curtailment compensation from RUC redispatch. Conversely, generators in persistently CON regions such as Zone 1 can anticipate being called upon beyond their day-ahead schedule through RUC. This gives rise to an incentive to bid high in the day-ahead market, since CON compensation is based on their offered price.

\begin{table}[htbp]
    \centering
    \renewcommand{\arraystretch}{1.4}
    \caption{Simulation Result of (Fall,Daytime) and (Spring,Daytime)}
    \resizebox{\textwidth}{!}{%
    \begin{tabular}{c|c|c|c|c}
        \hline
        \hline
        \multirow{2}{*}{\textbf{Province}} & \multicolumn{2}{c}{\textbf{(Fall,Daytime)}} & \multicolumn{2}{|c}{\textbf{(Summer,Evening)}}\\ 
        \cline{2-5} 
                              & Constrained-on  & Constrained-off & Constrained-on  & Constrained-off \\ 
        \hline
        \makecell{Zone\,1 \\[-0.4ex] {\scriptsize (Metropolitan)}}    & 1,940     & 1,391    & 476     & 0    \\
        \hline
        \makecell{Zone\,2 \\[-0.4ex] {\scriptsize (Chungcheong)}}          & 671       & 158      & 653     & 0    \\
        \hline
        \makecell{Zone\,3 \\[-0.4ex] {\scriptsize (Jeolla)}}           & -         & -        & 0     & 1,172     \\
        \hline
        \makecell{Zone\,4 \\[-0.4ex] {\scriptsize (Gyeongsang)}}           & 823       & 2,784    & 984     & 2,238   \\
        \hline
        \makecell{Zone\,5 \\[-0.4ex] {\scriptsize (Gangwon)}}              & 1,694     & 948      & 1,533   & 221     \\
        \hline
        \hline
    \end{tabular}}
    \label{tab:Redispatch result1}
\end{table}

Beyond the DAUC-RUC misalignment, both models share a common limitation: stability constraints are formulated as generator-side limits rather than transmission constraints. To satisfy transient and voltage stability criteria, the Korean power system relies on generator-side representations such as individual output caps, grouped generation constraints, and restrictions on the total number of committed units. 

However, this indirect representation cannot capture the economic value of network constraints, as generator-side limits do not produce shadow prices that can be reflected in market-clearing prices. Consequently, congestion costs that should emerge as regional price differences remain obscured, posing an obstacle to the effective implementation of locational pricing. 

\begin{figure}
    \centering
	\includegraphics[scale = 0.4]{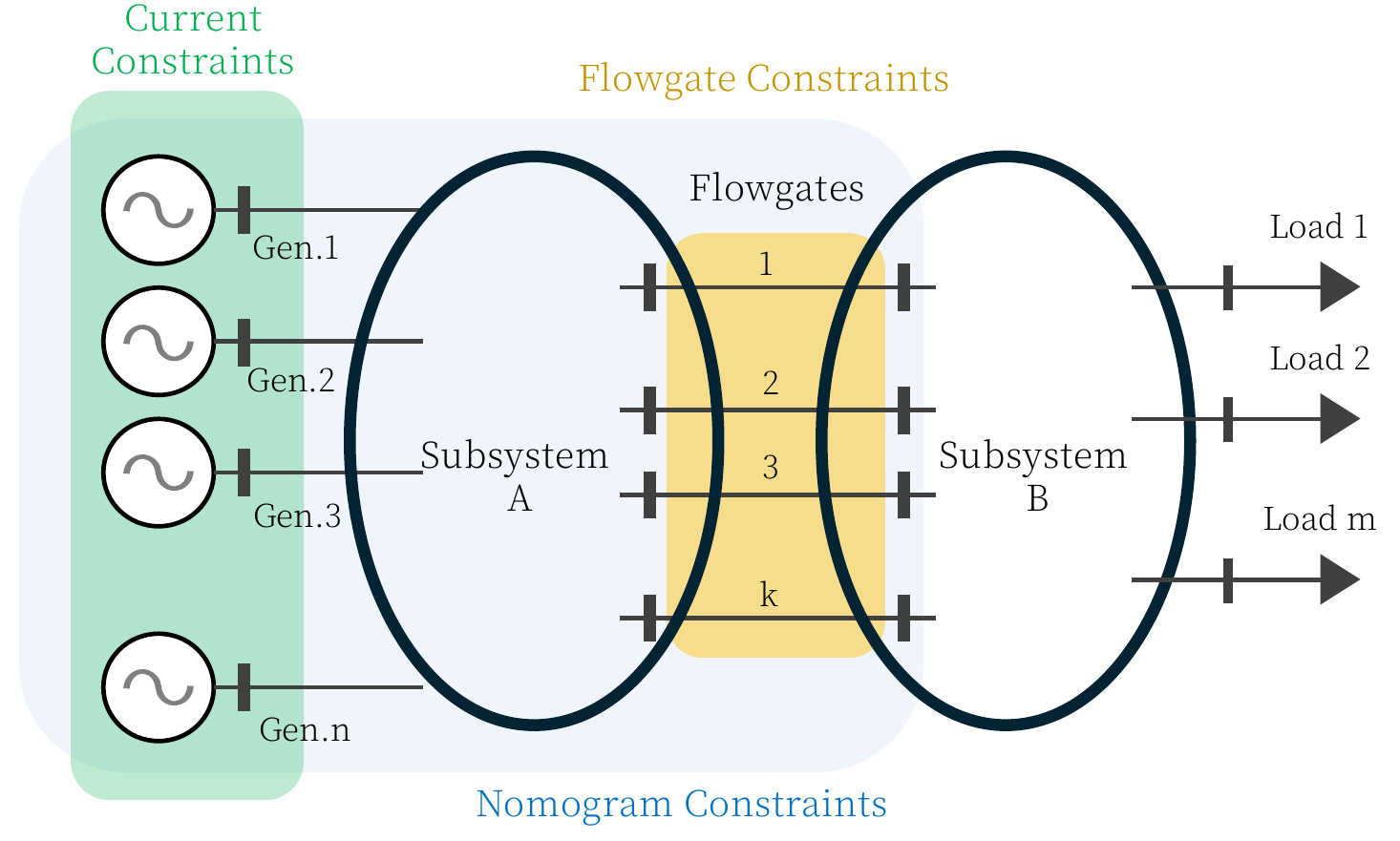}
	\caption{Nomogram and Flowgate Constraints}
	\label{fig:Constraints Illustration}
	\vspace{-1mm}
\end{figure}

To this end, stability constraints should be reformulated using transmission-side representations such as flowgate and nomogram constraints.Fig.~\ref{fig:Constraints Illustration} illustrates how these two approaches represent transmission constraints differently. Flowgate constraints limit active power flow across an interface, represented by the set of lines between subsystem A and B, using sensitivity factors of identified lines, while nomogram constraints define complex operating limits across combinations of operating variables for the entire subsystem. Critically, both can be incorporated directly into market-clearing optimization models, with their shadow prices captured as congestion prices. This approach has been used in MISO and CAISO, where shadow prices of binding flowgate and nomogram constraints are reflected in market-clearing prices. 

\subsubsection{Shift to Price-Based Offers} \label{subsubsec:PBP}
The current CBP mechanism relies on static cost parameters estimated by the system operator, which fail to capture the dynamic reality of generator operations. This deficiency is clearly illustrated in Table \ref{tab:Comparision}, which contrasts the cost components reflected in the price bidding of PJM with those used in Korea. The assessment excludes generator-specific conditions such as efficiency degradation, recognizes only hot start-up costs, unlike PJM which considers hot, warm, and cold starts, and explicitly excludes variable operation and maintenance (VOM) costs\footnote{VOM is a critical component of bidding strategies, precisely because it varies significantly by generator unit and is most accurately determined by the operators themselves. In many U.S. ISOs, the inclusion of VOM in a cost calculation enables the generator to ensure cost recovery through the wholesale market}. Ultimately, these exclusions mean that the assessed costs systematically fail to reflect the true economic costs of generation. 

The CBP was introduced as a pragmatic response to prevailing market constraints at the time. Since major generators were KEPCO subsidiaries, a substantial share of transactions was expected to occur within the same corporate group, necessitating limitations on bidding autonomy. Substituting competitive bidding with objective cost assessments effectively mitigated generator market power, prevented collusive activities, and contributed to price stabilization. However, the power system has undergone significant changes since then, which have exposed the inherent limitations of this approach more clearly. Renewable energy, demand response, and storage resources, which have near-zero variable costs and were virtually absent at the time of CBP adoption, now constitute a growing share of the generation mix, rendering administrative cost assessment increasingly difficult. Furthermore, as LNG fuel procurement has diversified beyond the average tariff supplied by KOGAS\footnote{The average tariff system applies a price averaged across all LNG import contracts held by KOGAS, regardless of the actual procurement cost of individual contracts.}, with direct imports by private generators increasing significantly, the standardized cost assessment framework is no longer able to accurately reflect the actual fuel costs incurred by individual generators. 

Price-based bidding has established itself as the dominant standard for wholesale electricity markets worldwide. Generators submit supply offer curves consisting of price-quantity pairs, a model widely adopted across most U.S. ISOs (\textit{e.g.}, PJM, CAISO, MISO, ISO-NE), New Zealand, and Colombia \citep{munoz2018economic}. The CBP approach remains limited to a small number of jurisdictions, notably Korea and parts of Latin America, such as Chile and Peru \citep{akcura2024global}. Under price-based bidding, generators can adjust their offers in response to changing market conditions, allowing prices to capture the true marginal value of resources and serve as reliable investment signals. 

However, implementing price-based bidding within a market with existing design flaws risks exacerbating rather than resolving inefficiencies. As demonstrated in Section \ref{subsubsec:LMP} and \ref{subsubsec:alignment}, both the uniform pricing and the DAUC-RUC misalignment would create systematic incentives for strategic bidding if price-based bidding were introduced without first addressing these underlying issues. Therefore, while the transition to price-based bidding is an essential long-term objective, rectifying the fundamental design flaws is a critical prerequisite. 

\begin{table}[htbp]
    \centering
    \renewcommand{\arraystretch}{1.4}
    \caption{Comparison of PJM and Korea}
    \begin{tabular}{c|c|c}
        \hline
        \hline
        \textbf{Variable Cost Properties} & 
        \textbf{PJM} & 
        \textbf{Korea} \\
        \hline
        \multirow{2}{*}[-0.6ex]{Fuel Cost} & $\bigcirc$ & $\bigcirc$ \\
        \cline{2-3}
        & \makecell{Determination by \\the generator itself} & \makecell{Monthly heat value \\ unit price} \\
        \hline
        Environmental Cost & $\bigcirc$ & $\bigcirc$ \\
        \hline
        Emission Cost & $\bigcirc$ & $\bigcirc$ \\
        \hline
        \makecell{Degradation due to aging \\ \& Deviation in \\ performance test} & $\bigcirc$ & $\times$ \\
        \hline
        No-load Cost & $\bigcirc$ & $\bigcirc$ \\
        \hline
        Start-up Cost & \makecell{$\bigcirc$ \\ (Hot/Warm/Cold)}& $\bigcirc$ \\
        \hline
        VOM & $\bigcirc$ & $\times$ \\
        \hline
        \hline
    \end{tabular}
    \label{tab:Comparision}
\end{table}

\subsection{Implementation Pathway}
\begin{figure}
    \centering
	\includegraphics[scale = 0.25]{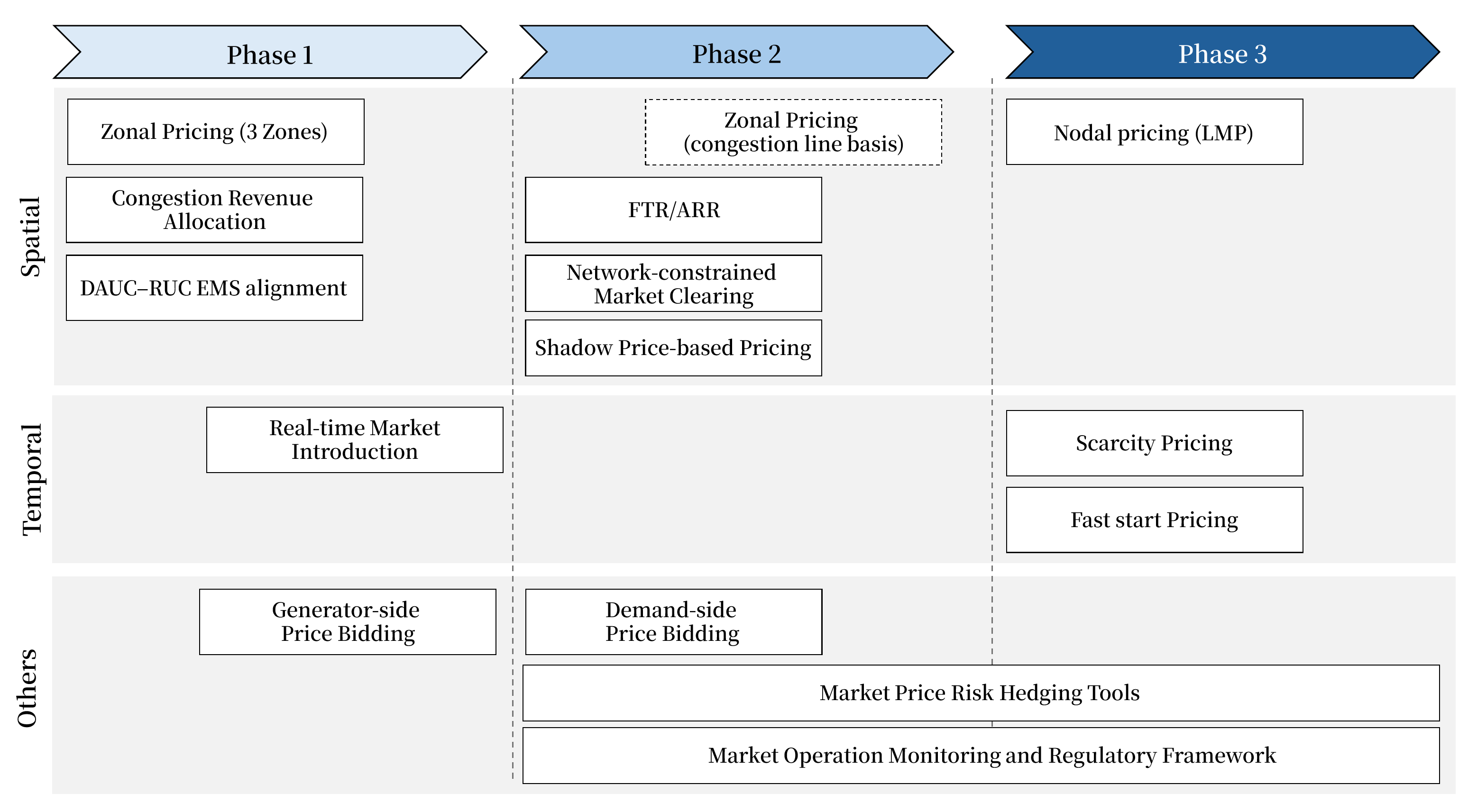}
	\vspace{-2mm}
	\caption{Roadmap for Wholesale Electricity Market Reform in Korea}
	\label{fig:roadmap}
	\vspace{-1mm}
\end{figure}

Drawing on the discussed in the preceding sections, this section presents a three-phase implementation roadmap for improving the price formation function of the Korean electricity market, as illustrated in Fig.~\ref{fig:roadmap}. The roadmap is structured as a sequentially dependent reform trajectory, in which each phase establishes the institutional and technical preconditions upon which the subsequent phase rests. Individual measures within each phase are likewise mutually interdependent. A pragmatic alternative is to prioritize and sequence measures according to their relative importance and urgency. It should be recognized, however, that because the measures are interlinked, some degree of market inefficiency may be unavoidable until preceding measures are fully in place.

\subsubsection{Phase 1: Establishing Market Structural Foundations}
To improve the spatial dimension of price signals, Phase 1 introduces zonal pricing using the MA and NMA boundary already used in DAUC as the initial step. This approach reflects an existing operational reality rather than imposing an entirely new institutional boundary, preserving stakeholder familiarity. Zonal pricing, however, cannot stand alone. The congestion rents arising from locational price differentials constitute market surplus and must be returned to market participants through a concurrently established allocation mechanism. Given implementation constraints in the Korean context, a pragmatic interim approach is to begin with direct allocation to designated participants, with the allocation framework progressively refined as the market matures. 

Alongside these pricing reforms, DAUC–RUC alignment must be established as a precondition for the remaining Phase 1 measures. As demonstrated in Section \ref{subsubsec:alignment}, the two engines operate under fundamentally different constraint sets, producing redispatch that is asymmetric across regions rather than random. In the absence of a real-time market, RUC currently serves as the closest approximation to real-time dispatch in the Korean system. The introduction of a real-time market will require its role to be redefined — narrowed to a residual reliability tool addressing only what the day-ahead market leaves unresolved, as in PJM and ISO-NE. For this narrowing to be operationally coherent, the constraint divergence between DAUC and RUC should first be reduced; as long as the two engines operate under materially different constraint sets, it would be difficult to define the scope of RUC's residual role in a principled way. The same alignment is a prerequisite for price-based bidding: the predictable asymmetry of current redispatch patterns would generate perverse strategic bidding incentives if price-based offers were introduced under the existing structure.

To improve the temporal dimension of price signals, Phase 1 introduces a real-time market operating at five- or fifteen-minute intervals. By providing settlement prices that correspond to actual generation quantities, a real-time market reflects the temporal value of electricity more accurately than any alternative structure, and at a time when flexible resource investment is critically needed, real-time price signals are essential for enabling market-driven investment decisions. 

Finally, the cost-based bidding regime is replaced with a price-based framework in which generators submit offers reflecting their own marginal costs. This allows generation scheduling and dispatch to align more closely with actual cost structures, reducing system operation costs while enabling market prices to reflect marginal values with greater precision. Since competitive price formation in any subsequent phase presupposes price-based offers rather than cost reports, deferring this transition would undermine the feasibility of the reform as a whole.

\subsubsection{Phase 2: Refining Price Formation and Expanding Participation}
On the spatial dimension, a further refinement to congestion line-based zonal pricing is possible but not essential at this stage. While a more granular zonal structure would sharpen locational price signals, the institutional and technical effort required to implement it as an intermediate step before nodal pricing risks being disproportionate to the benefit. What Phase 2 does require, however, is the completion of the congestion revenue allocation framework. Whereas Phase 1 adopts a direct allocation approach as an interim measure, Phase 2 transitions toward competitive auction mechanisms such as FTRs, through which market participants acquire congestion rent entitlements through market competition — the arrangement most consistent with efficient market design principles.

The more consequential pricing reform in Phase 2 concerns the shift from SMP-based to shadow price-based pricing. The zonal pricing introduced in Phase 1 remains compatible with the existing SMP screening approach, but finer locational price signals require that settlement prices be derived from the dual variables of the market clearing optimization. For shadow prices to accurately reflect transmission constraints, however, the stability constraints currently embedded in system operation must be reformulated as explicit network constraints within the clearing model — so that congestion costs are transparently priced rather than administratively absorbed.

With the real-time market now operational and price volatility correspondingly elevated, Phase 2 extends price bidding to the demand side. Permitting demand response resources to adjust consumption in response to real-time price signals would introduce the supply-demand symmetry that meaningful price formation requires. The increase in price volatility also necessitates the parallel development of financial hedging instruments such as electricity futures, providing market participants with the means to manage price risk exposure. Finally, as the market grows in complexity across all these dimensions, independent market monitoring becomes indispensable. Surveillance of bidding behavior, detection of price anomalies, and ex-post settlement verification must be institutionalized at this stage, as these functions underpin the integrity of the fully developed market of Phase 3.

\subsubsection{Phase 3: Toward Full Marginal Cost Transparency}
Phase 3 completes the reform through the introduction of nodal pricing, which represents the finest achievable spatial resolution of locational price signals — each node reflecting the precise marginal value of resources at that location. This eliminates the spatial aggregation distortions that persist under zonal pricing and provides market prices most closely aligned with actual generation, made technically feasible by the network-constrained market clearing and shadow price-based pricing established in Phase 2.

By the time Phase 3 is reached, demand for flexible resources is likely to be considerably greater than at present. While real-time market prices already provide investment signals for flexibility, additional measures may be warranted depending on market conditions. If fast-start units such as gas peakers play a particularly critical role in system operation, extending price-setting eligibility to these resources would strengthen the price formation function and better reflect the marginal value of fast-response capacity. Similarly, if substantial renewable energy expansion has suppressed average market prices to the point where investment incentives for new flexible capacity are insufficient — despite growing reserve requirements — applying an Operating Reserve Demand Curve (ORDC) should be considered. By pricing reserve scarcity directly into the energy market, an ORDC provides the investment signals necessary to attract new capacity and support long-term system reliability without relying on administrative procurement.

\begin{landscape}
\begin{figure}
    \centering
	\includegraphics[scale = 0.3]{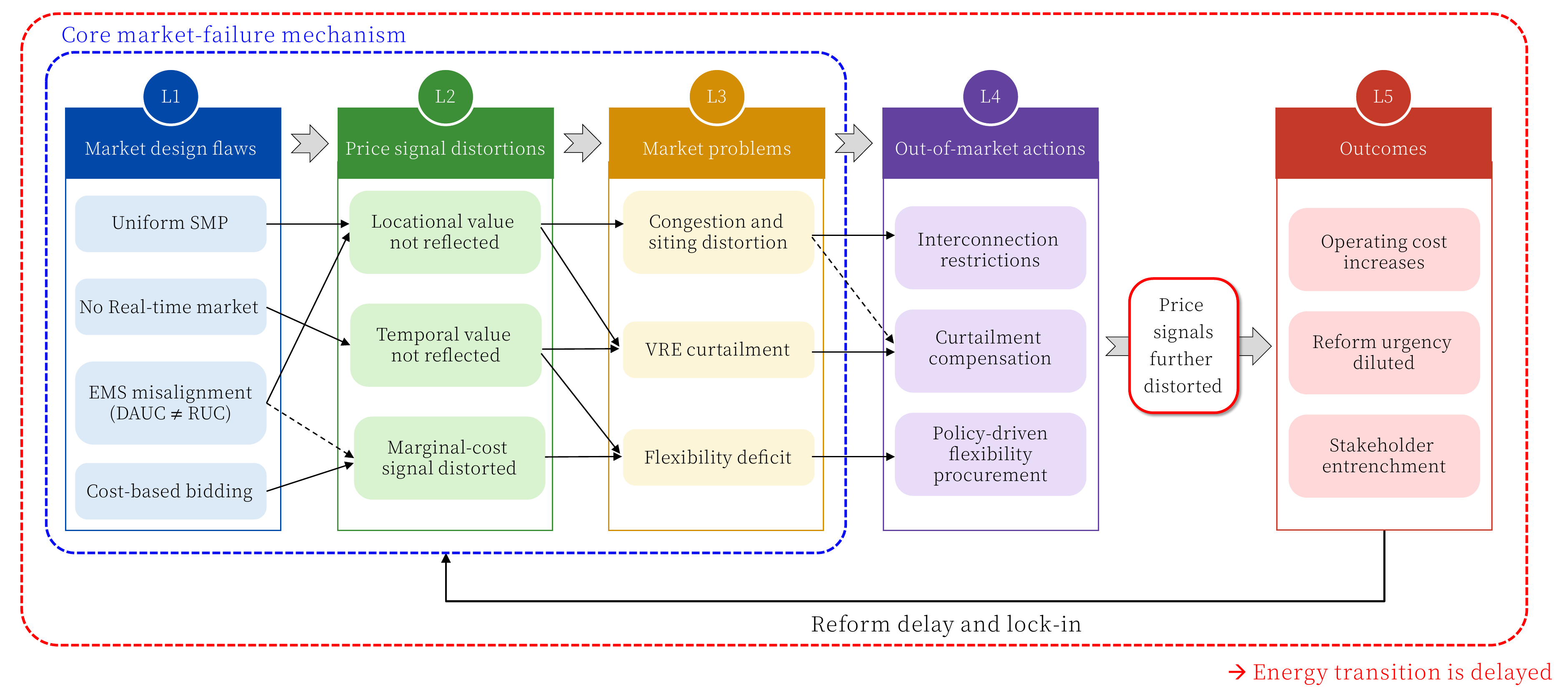}
	\caption{Korean Market Failure Chain}
	\label{fig:failure}
	\vspace{-1mm}
\end{figure}
\end{landscape}

\section{Discussion and Conclusion} \label{Sec5:Discussion}
\subsection{Breaking the Policy Vicious Cycle}
Fig.~\ref{fig:failure} shows the failure chain through which Korea's electricity market design flaws compound into systemic outcomes. Market design flaws at L1 produce distorted price signals at L2, which generate the operational problems at L3: congestion, curtailment, and a flexibility deficit. The critical dynamic, however, lies in what follows. The administrative interventions deployed at L4 to manage these problems do not address their root cause; instead, they further distort the already-compromised price signals, driving the outcomes at L5 that will be discussed below.

The outcomes at L5 close this loop across three dimensions. First, operating costs rise directly: without accurate locational and temporal price signals, dispatch deviates systematically from efficient allocations, while administrative interventions add their own costs through curtailment compensation and centralized BESS procurement. Second, these same interventions dilute the perceived urgency of reform. By providing apparent operational stability, they make the underlying price signal failure less visible to market participants and regulators alike, weakening the drive for structural change. Third, and most consequentially, some participants come to benefit from the distorted status quo: generators in persistently curtailed regions receive compensation that would disappear under market-clearing prices, while BESS operators with long-term guaranteed contracts have little incentive to support a shift to market-based procurement. This entrenchment progressively erodes the political feasibility of reform and locks the market further into the failure chain.

The longer administrative interventions persist, the harder reform becomes. As renewable penetration continues to rise, the operational stress generated by these distortions will only compound, while the institutional capacity to respond through market mechanisms erodes in parallel. Administrative management of these failures is at best a temporary stopgap. Restoring the price formation function of the market is not one option among many — it is the only path through which Korea can achieve both an efficient energy system and a credible energy transition.

\subsection{Limitations and Future Work}
This research focuses on restoring price signals within the wholesale electricity market. It does not address the interaction between the energy market and either the capacity market or the ancillary service market, which should be examined to ensure coherent signals across market segments. 

The proposed reforms are based on standard market architecture found in leading international electricity markets. While this provides a robust benchmark for identifying flaws within the existing regime, the adaptation of these universal mechanisms to Korea's unique grid characteristics and market governance remains a critical direction for future research. Although this research identifies the direction and rationale for reform, the detailed design of individual market mechanisms remains to be developed. Moreover, as market complexity increases with reform, robust market monitoring and mitigation mechanisms to prevent the abuse of market power, along with financial instruments to manage price volatility, must be designed.

\subsection{Conclusion}
This study demonstrates that the price formation in the Korean electricity market is structurally flawed, and that these distortions extend beyond operational inefficiency to fundamentally undermine national energy transition goals. Our analysis reveals that the current wholesale market, characterized by inaccurate network representation, fundamental misalignment between market and system operations, and a rigid CBP framework, does not reflect the spatial and temporal value of electricity. This flawed signaling compromises resource allocation and hinders the large-scale integration of renewable energy and enhancement of grid flexibility. 

Addressing these challenges requires integrated, market-based reform. We propose four complementary reforms\textemdash introducing locational marginal pricing, establishing a real-time market, unifying market and system operations, and transitioning to a price-based bidding system\textemdash that collectively restore reliable price signals. The synergy of these reforms, rather than isolated technical fixes, aligns the profit-maximizing incentives of individual participants with system-wide efficiency. This provides the foundation for short-run operational efficiency while allowing price signals to guide long-term investment toward national energy goals.


\appendix
\section{Four-node Example for Locational Pricing} \label{app1}
The failure to explicitly incorporate transmission constraints into the pricing model fundamentally distorts price signals, necessitating significant uplift payments to compensate for the mismatch between dispatch and pricing. To illustrate how such inefficiencies emerge when transmission value is not properly reflected in market prices, this section employs a simplified four-bus test case to compare market outcomes under different pricing mechanisms.

As shown in Fig.~\ref{fig:four-node example}, the test system comprises four buses: three in Zone 1 and one in Zone 2. Each bus is equipped with a generator, while the entire demand is concentrated at Bus 4. The marginal cost of each generator is shown in the Fig.~\ref{fig:four-node example}, with the willingness to pay for demand set at \$100/MWh. The transmission lines are assumed to be lossless, and the inter-zonal tie line has a transfer capacity of 500\,MW.

For comparison, this case study is analyzed using two market clearing models:a comprehensive nodal pricing model and a simplified zonal pricing model. The nodal model is theoretically more accurate as it incorporates complete network information into the market clearing process, whereas the zonal model only considers constraints on inter-zonal lines\footnote{The zonal pricing discussed in this section refers to the ATC (Available Transfer Capability)-based method.}. In zonal pricing, since the limits of intra-zonal lines are not accounted for in the market clearing, system operators must rely on measures such as tightening interconnection line capacity or imposing generation constraints on specific units to prevent congestion on these lines during real-time operations.

Tables \ref{tab:Nodal_Result(Quantites)} and \ref{tab:Nodal_Result(Price)} present the generation dispatch and market prices derived from the optimal power flow analysis under nodal pricing. To meet the total system demand of 800\,MW, the dispatch levels for generators $P_1$,$P_2$,$P_3$, and $P_4$ are determined to be 175\,MW, 100\,MW, 225\,MW, and 300\,MW, respectively. Consequently, the total generation cost is calculated to be \$26,750. Since this dispatch fully incorporates transmission line capacity limits, it achieves the lowest possible generation cost within the set of feasible solutions. Notably, the market price was highest at Bus 4, where demand is concentrated. Despite the identical marginal costs of generators at Buses 1 and 2, the system assigns different marginal values to electricity at each location due to transmission congestion, causing the market prices to diverge to \$10/MWh at Bus 1 and \$25/MWh at Bus 2. Social welfare, defined as the sum of consumer surplus, generator surplus, and congestion rent, amounts to \$53,250 under the nodal pricing. 

\begin{table}[htbp]
    \centering
    \renewcommand{\arraystretch}{1.4}
    \caption{Nodal Pricing Market Clearing Quantity}
    \begin{tabular}{c|c}
        \hline
        \hline
        \textbf{Generator} & \textbf{Market Clearing Quantity } \\
        \hline
        $P_1^{*}$ & 175~MW\\
        \hline
        $P_2^{*}$ & 100~MW\\
        \hline
        $P_3^{*}$ & 225~MW\\
        \hline
        $P_4^{*}$ & 300~MW\\
        \Xhline{1.5pt}  
        Total Generation Cost & \$26,750 \\
        \hline
        \hline
    \end{tabular}
    \label{tab:Nodal_Result(Quantites)}
\end{table}

\begin{table}[htbp]
    \centering
    \renewcommand{\arraystretch}{1.4}
    \caption{Nodal Market Prices and Social Welfare Summary}
    \begin{tabular}{c|c|c|c|c|c|c|c}
        \hline
        \hline
        \multicolumn{4}{c|}{\makecell{Nodal Price\\(\$/MWh)}} & 
        \multirow{2}{*}{\makecell{Generator\\Revenue}} & 
        \multirow{2}{*}{\makecell{Consumer\\Payment}} &
        \multirow{2}{*}{\makecell{Congestion\\Rent}} & 
        \multirow{2}{*}{\makecell{Social\\Surplus}} \\
        \cline{1-4}
        $\lambda_1^*$ & $\lambda_2^*$ & $\lambda_3^*$ & $\lambda_4^*$ & & & & \\
        \hline
        10 & 25 & 40 & 50 & \$28,250 & \$40,000 & \$11,750 & \$53,250 \\
        \hline
        \hline
    \end{tabular}
    \label{tab:Nodal_Result(Price)}
\end{table}

Next, we examine the market clearing results under the zonal pricing mechanism, as presented in Tables \ref{tab:Zonal_Result(Quantities)} and \ref{tab:Zonal_Result(Price)}. Since intra-zonal transmission limits are disregarded, generation within each zone is dispatched in ascending order of marginal cost. As a result, zonal prices are determined by the incremental generation cost associated with a 1\,MW increase in demand within each zone, resulting in prices of \$40/MWh for Zone\,1 and \$50/MWh for Zone\,2. However, the resulting dispatch of 200\,MW, 100\,MW, and 200\,MW from generators 1, 2, and 3 in Zone 1 violates intra-zonal transmission constraints. Specifically, the power flow on line 1-3 reaches 167\,MW, exceeding its line capacity of 150\,MW. Thus, the generation dispatch schedule derived from this market clearing is physically infeasible. To resolve this issue under zonal pricing, the system operator must resort to out-of-market actions, such as reducing inter-zonal transmission capacity or directly constraining specific generators, for instance, by imposing upper and lower bounds on their generation.

\begin{table}[htbp]
    \centering
    \renewcommand{\arraystretch}{1.4}
    \caption{Zonal Pricing Market Clearing Quantity}
    \begin{tabular}{c|c}
        \hline
        \hline
        \textbf{Generator} & \textbf{Market Clearing Quantity} \\
        \hline
        $P_1^{*}$ & 200~MW \\
        \hline
        $P_2^{*}$ & 100~MW \\
        \hline
        $P_3^{*}$ & 200~MW \\
        \hline
        $P_4^{*}$ & 300~MW \\
        \hline
        \hline
    \end{tabular}
    \label{tab:Zonal_Result(Quantities)}
\end{table}

\begin{table}[htbp]
    \centering
    \renewcommand{\arraystretch}{1.4}
    \caption{Zonal Market Prices and Welfare Summary}
    \begin{tabular}{c|c|c|c}
        \hline
        \hline
        \multicolumn{2}{c|}{\makecell{Zonal Price\\(\$/MWh)}} & 
        \multirow{2}{*}{\makecell{Generator\\Revenue}} & 
        \multirow{2}{*}{\makecell{Consumer\\Payment}} \\
        \cline{1-2}
        Zone 1 & Zone 2 &  \\
        \hline
        40 & 50 & \$35,000 & \makecell{\$40,000\\ (800MW*\$50)} \\
        \hline
        \hline
    \end{tabular}
    \label{tab:Zonal_Result(Price)}
\end{table}

\begin{table}[htbp]
    \centering
    \renewcommand{\arraystretch}{1.4}
    \caption{Zonal Pricing Market Clearing Quantity (Case (1): Inter-zonal Line Limit)}
    \begin{tabular}{c|c}
        \hline
        \hline
        \textbf{Generator} & {\makecell{\textbf{Market Clearing Quantity} \\ Inter-zonal transmission line capacity:270MW}} \\
        \hline
        $P_1^{*}$ & 180~MW \\
        \hline
        $P_2^{*}$ & 90~MW \\
        \hline
        $P_3^{*}$ & 0~MW \\
        \hline
        $P_4^{*}$ & 530~MW \\
        \Xhline{1.5pt}  
        Total Generation Cost & \$29,200 \\
        \hline
        \hline
    \end{tabular}
    \label{tab:Zonal_Result2(Quantities)}
\end{table}

\begin{table}[htbp]
    \centering
    \renewcommand{\arraystretch}{1.4}
    \caption{Zonal Market Prices and Welfare Summary (Case (1): Inter-zonal Line Limit)}
    \begin{tabular}{c|c|c|c|c|c}
        \hline
        \hline
        \multicolumn{2}{c|}{\makecell{Zonal Price\\(\$/MWh)}} & 
        \multirow{2}{*}{\makecell{Generator\\Revenue}} & 
        \multirow{2}{*}{\makecell{Consumer\\Payment}} &
        \multirow{2}{*}{\makecell{Congestion\\Rent}} & 
        \multirow{2}{*}{\makecell{Social\\Surplus}} \\
        \cline{1-2}
        Zone 1 & Zone 2 & & &  \\
        \hline
        10 & 50 & \$29,200 & \$40,000 & \$10,800 & \$50,800 \\
        \hline
        \hline
    \end{tabular}
    \label{tab:Zonal_Result2(Price)}
\end{table}

First, we consider a method where the inter-zonal transmission limit is tightened from 500\,MW to 270\,MW to mitigate intra-zonal congestion. As a result, the revised dispatch levels for generators 1,2, and 3 in Zone\,1 are 180\,MW, 90\,MW, and 0\,MW, respectively, as shown in Table \ref{tab:Zonal_Result2(Quantities)}. While this generation dispatch is now feasible, constraining the inter-zonal line limit forces the system to use transmission capacity inefficiently. This, in turn raises total operational costs by \$2,450/MWh compared to the nodal pricing case. All generators within each zone are subject to a uniform market price, set at \$10/MWh for Zone 1 and \$50/MWh for Zone 2. In Zone 1, the market price aligns with the variable cost of the marginal generator, yielding zero profit for all dispatched generators. This loss of surplus is particularly pronounced for generator\,2. Under nodal pricing, generator\,2 earned a profit of \$1,500; however, under this zonal scheme, its profit is entirely eliminated. In summary, restricting inter-zonal transmission capacity to alleviate intra-zonal congestion incurs higher system operating costs and decreases generator surplus. Consequently, social welfare under this zonal approach diminishes by \$2,450 relative to the nodal pricing outcome. 

\begin{table}[htbp]
    \centering
    \renewcommand{\arraystretch}{1.4}
    \caption{Zonal Pricing Market Clearing Quantity (Case (2): Generation Limit)}
    \begin{tabular}{c|c}
        \hline
        \hline
        \textbf{Generator} & \textbf{Market Clearing Quantity} \\
        \hline
        $P_1^{*}$ & 175~MW \\
        \hline
        $P_2^{*}$ & 100~MW \\
        \hline
        $P_3^{*}$ (Minimum bound) & 225~MW \\
        \hline
        $P_4^{*}$ & 300~MW \\
        \Xhline{1.5pt} 
        Total Generation Cost & \$26,750 \\
        \hline
        \hline
    \end{tabular}
    \label{tab:Zonal_Result3(Quantities)}
\end{table}

Next, we examine how imposing generation limits on specific units can alleviate intra-zonal congestion. In real-world power systems, it is computationally challenging to derive precise generation constraints that perfectly represent the effects of physical transmission limits. However, for this simplified example, we assume that the ideal generation limits are identified through an iterative process or detailed power flow analysis. By imposing a minimum generation constraint of 225\,MW on generator 3, the market clears with dispatch levels for generators 1 and 2 at 175\,MW and 100\,MW, respectively. As shown in Table \ref{tab:Zonal_Result3(Quantities)}, this resulting dispatch is identical to the nodal pricing outcome. 

However, the resulting market prices differ from the outcomes of the previous nodal and zonal pricing cases. As Generator\,3 is binding at its minimum output constraint, it cannot set the market price. Instead, the market price in Zone 1 is determined by the marginal cost of Generator 1, which is \$10/MWh, while the market price in Zone 2 settles at \$50/MWh, consistent with earlier examples. 
Generator 3, with a marginal cost of \$40/MWhm is forced to operate at a loss given the clearing price of \$10/MWh. Thus, the market operator is required to provide an uplift payment of \$6,750 to ensure cost recovery. Table \ref{tab:Zonal_Result3(Price)} presents the generator revenues, including both market settlements and uplift payments. Consequently, total consumer payments amount to \$46,750, which includes the market payment of \$40,000 and an additional uplift payment of \$6,750. As a result, the minimum output requirement on Generator 3 forces the Zone 1 market price to a level that fails to reflect the true marginal value of electricity, necessitating substantial out-of-market payments. 

This example demonstrates that when transmission constraints are explicitly reflected in market clearing, the economic value of congestion is more accurately captured in market prices, which improves price signals and reduces unnecessary uplift payments.

\begin{table}[htbp]
    \centering
    \renewcommand{\arraystretch}{1.4}
    \caption{Zonal Market Prices and Welfare Summary (Case (2): Generation Limit)}
    \begin{tabular}{c|c|c|c|c|c}
        \hline
        \hline
        \multicolumn{2}{c|}{\makecell{Zonal Price\\(\$/MWh)}} & 
        \multirow{2}{*}{\makecell{Generator\\Revenue}} & 
        \multirow{2}{*}{\makecell{Consumer\\Payment}} &
        \multirow{2}{*}{\makecell{Congestion\\Rent}} & 
        \multirow{2}{*}{\makecell{Social\\Surplus}} \\
        \cline{1-2}
        Zone 1 & Zone 2 & & &  \\
        \hline
        10 & 50 & \makecell{\$26,750 \\ (\$20,000+ \\ \$6,750)} & \makecell{\$46,750 \\ (\$40,000+ \\ \$6,750)} & \$20,000 & \$53,250 \\
        \hline
        \hline
    \end{tabular}
    \label{tab:Zonal_Result3(Price)}
\end{table}

\section{Simulation Details} \label{app2}
We constructed a simulation model that replicates the operating environment of the actual Korean power system. Within this framework, the grid topology includes all buses and thousands of transmission lines rated at 154\,kV. The nodal load distribution was reconstructed from data in the Basic Plan for Long-term Electricity Supply and Demand. Table \ref{tab:Generator Summary} shows the generator type, total installed generator capacity, and the average fuel price by generator type. Renewable energy output and demand for each scenario was based on average historical data in 2023 \citep{data, KPX}.

\begin{table}[ht]
    \centering
    \caption{Criteria defining of each snapshots}
    \renewcommand{\arraystretch}{1.2}
    \begin{tabular}{c|c||c|c}
    \hline
    \hline
    Season & Months & Time & Hours \\ 
    \hline
    Spring & 3 - 5 & Morning & 05:00 $\sim$ 09:00 \\ 
    \hline
    Summer & 6 - 8 & Daytime & 10:00 $\sim$ 16:00 \\ 
    \hline
    Fall & 9 - 11 & Evening & 17:00 $\sim$ 22:00 \\ 
    \hline
    Winter & 12 - 1 & Night & 23:00 $\sim$ 04:00 \\ 
    \hline
    \hline
    \end{tabular}
    \label{tab:seasonal_time_division}
\end{table}

\begin{table}
    \centering
	\caption{Network Information about the Korean Power System}
        \renewcommand{\arraystretch}{1.2}
        \begin{tabular}{c|c|c}
        \noalign{\smallskip}\noalign{\smallskip}\hline\hline
        \multicolumn{2}{c|}{Network Components} & Number \\
        \hline
        \multirow{5}{*}{Bus} 
        & 765\,kV & $9$ \\
        \cline{2-3}
        & 345\,kV & $203$  \\
        \cline{2-3}
        & 154\,kV & $154$  \\
        \cline{2-3}
        & lower voltage & $2,642$  \\
        \cline{2-3}
        & Total & $4,240$ \\
        \hline
        \multirow{3}{*}{Machine} 
        & Synchronous Generator & $480$ \\
        \cline{2-3}
        & Renewables & $1,137$ \\
        \cline{2-3}
        & Total & $1,617$ \\
        \hline
        \multirow{1}{*}{Load} 
        & Total & $1,518$ \\
        \hline
        \multirow{4}{*}{Branch}
        & Transmission Line & $3,066$ \\
        \cline{2-3}
        & Transformer (2-Winding) & $2,238$ \\
        \cline{2-3}
        & Transformer (3-Winding) & $369$ \\
        \cline{2-3}
        & Total & $5,673$ \\ 
        \hline
        \hline
        \end{tabular}
    \label{table:SouthKR}
\end{table}

\begin{figure}[htbp]
    \centering
    \begin{subfigure}[b]{0.35\textwidth}
        \centering
        \includegraphics[width=\linewidth]{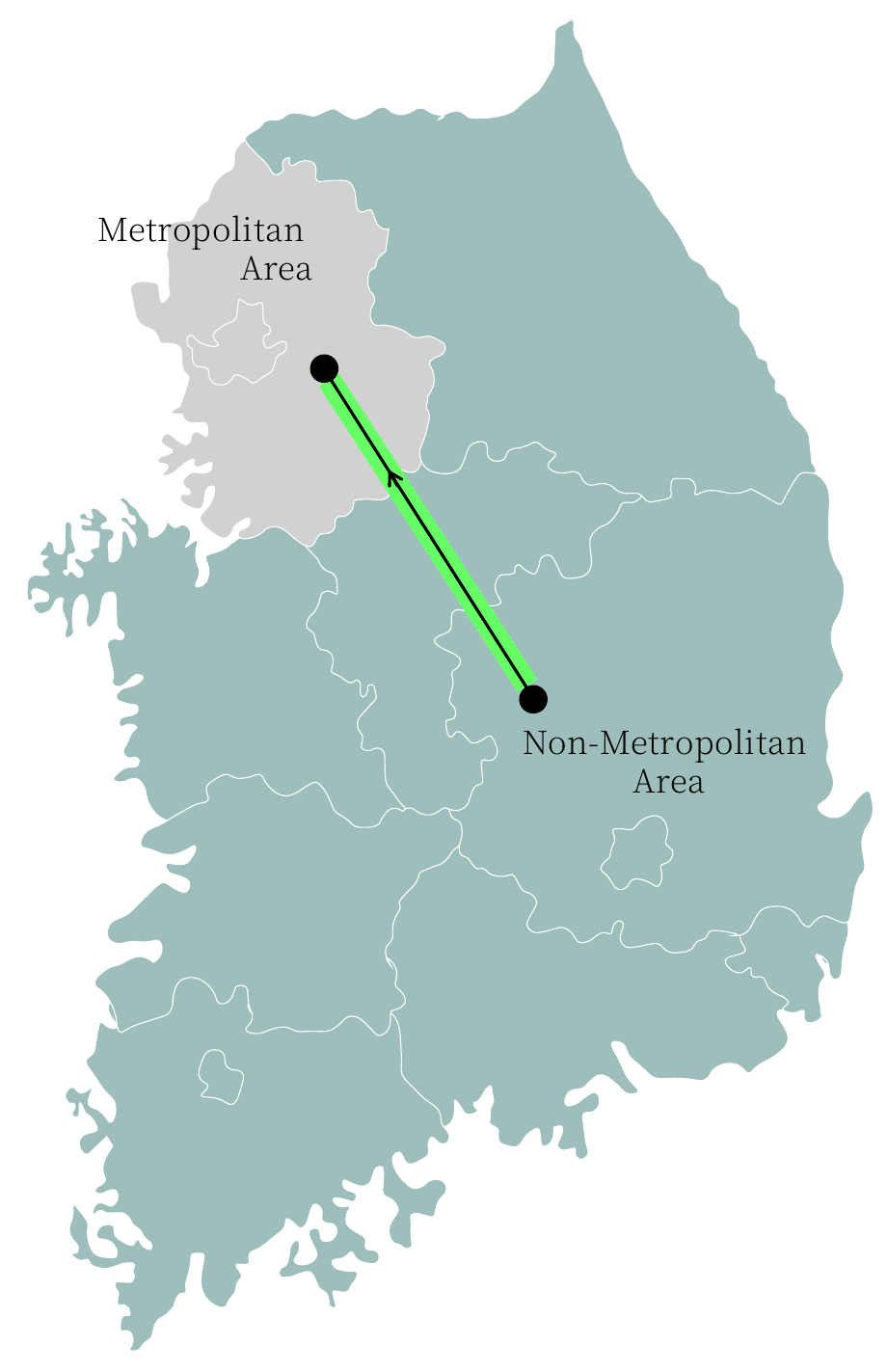}
        \caption{Zonal}
        \label{fig:ZonalMap}
    \end{subfigure}
    \begin{subfigure}[b]{0.36\textwidth}
        \centering
        \includegraphics[width=\linewidth]{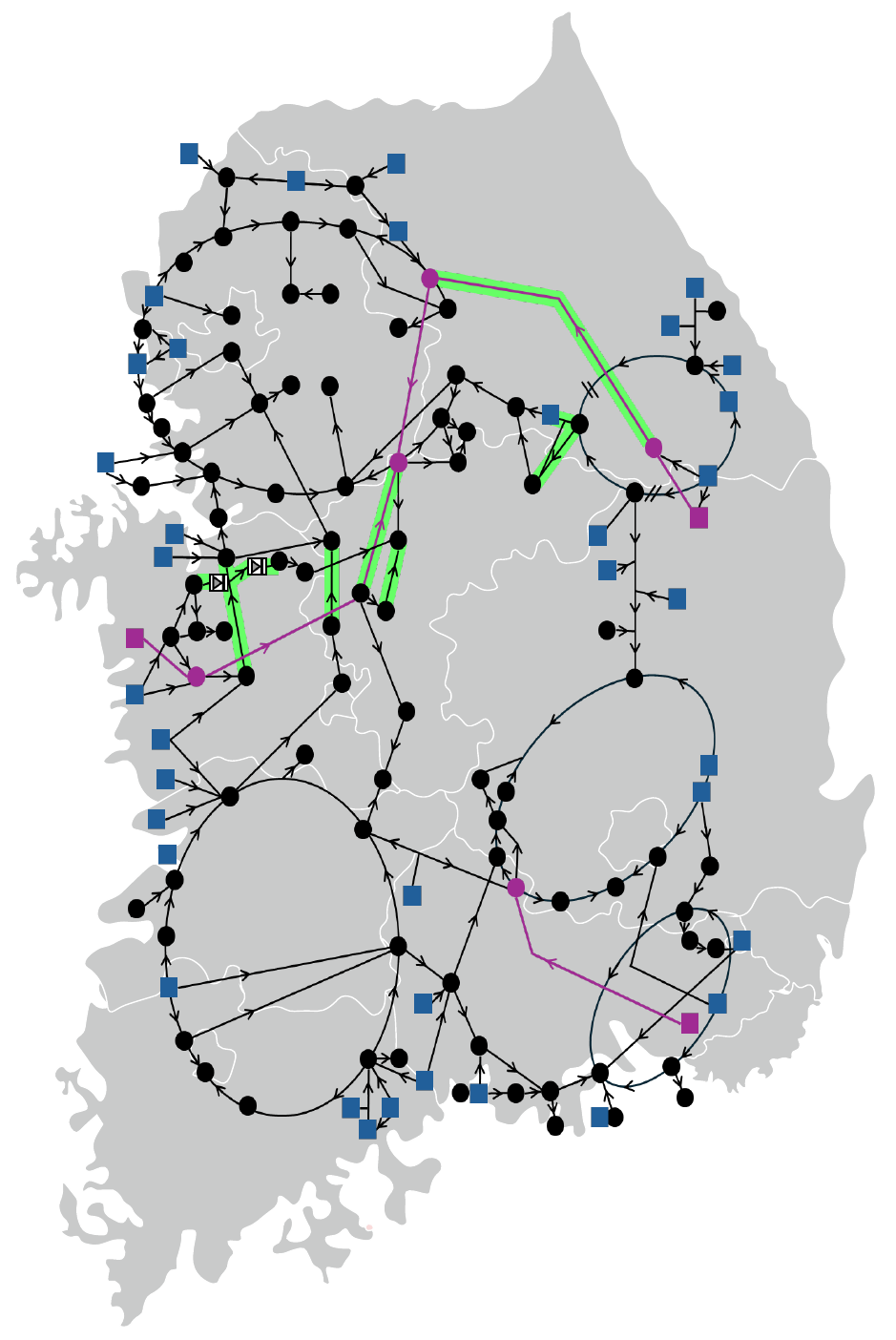}
        \caption{Nodal}
        \label{fig:NodalMap}
    \end{subfigure}

    \caption{The topology of Korean Power System Modeling}
    \label{fig:Map}
\end{figure}

\begin{table}[htbp]
    \centering
    \renewcommand{\arraystretch}{1.4}
    \caption{Summary of Generator Information}
    \begin{tabular}{c|c|c}
        \hline
        \hline
        \textbf{Category} & \makecell{\textbf{Installed Capacity} \\ \textbf{(MW)}} & \makecell{\textbf{Average Fuel} \\ \textbf{Price (\$/GJ)}}\\
        \hline
        Nuclear & 26,722 & 2.58\\
        \hline
        Coal & 37,299 & 37.49\\
        \hline
        Oil & 526 & 104.85\\
        \hline
        LNG & 51,725 & 89.62\\
        \hline
        Hydrogen & 1,630 & -\\
        \hline
        Pumped Storage & 4,700 & -\\
        \hline
        Wind & 6,253 & -\\
        \hline
        Solar PV & 27,871 & -\\
        \hline
        Others & 5,814 & -\\
        \hline
        \hline
    \end{tabular}
    \label{tab:Generator Summary}
\end{table}

\bibliographystyle{elsarticle-harv} 
\bibliography{Reference}

@article{MUNOZ2021111997,
title = {Electricity market design for low-carbon and flexible systems: Room for improvement in Chile},
journal = {Energy Policy},
volume = {148},
pages = {111997},
year = {2021},
issn = {0301-4215},
doi = {https://doi.org/10.1016/j.enpol.2020.111997},
author = {Francisco D. Muñoz and Carlos Suazo-Martínez and Eduardo Pereira and Rodrigo Moreno},
}

@article{NEWBERY2018695,
title = {Market design for a high-renewables European electricity system},
journal = {Renewable and Sustainable Energy Reviews},
volume = {91},
pages = {695-707},
year = {2018},
issn = {1364-0321},
doi = {https://doi.org/10.1016/j.rser.2018.04.025},
author = {David Newbery and Michael G. Pollitt and Robert A. Ritz and Wadim Strielkowski},
}

@article{LYNCH2021101312,
title = {Market design options for electricity markets with high variable renewable generation},
journal = {Utilities Policy},
volume = {73},
pages = {101312},
year = {2021},
issn = {0957-1787},
doi = {https://doi.org/10.1016/j.jup.2021.101312},
url = {https://www.sciencedirect.com/science/article/pii/S0957178721001454},
author = {Muireann Lynch and Genaro Longoria and John Curtis},
}

@article{CONEJO2018520,
title = {Rethinking restructured electricity market design: Lessons learned and future needs},
journal = {International Journal of Electrical Power \& Energy Systems},
volume = {98},
pages = {520-530},
year = {2018},
issn = {0142-0615},
doi = {https://doi.org/10.1016/j.ijepes.2017.12.014},
author = {Antonio J. Conejo and Ramteen Sioshansi},
}

@article{LESLIE2020106847,
title = {Designing electricity markets for high penetrations of zero or low marginal cost intermittent energy sources},
journal = {The Electricity Journal},
volume = {33},
number = {9},
pages = {106847},
year = {2020},
note = {Special Issue: The Future Electricity Market Summit},
issn = {1040-6190},
doi = {https://doi.org/10.1016/j.tej.2020.106847},
url = {https://www.sciencedirect.com/science/article/pii/S1040619020301391},
author = {Gordon W. Leslie and David I. Stern and Akshay Shanker and Michael T. Hogan},
}

@article{LI2019330,
title = {Using diverse market-based approaches to integrate renewable energy: Experiences from China},
journal = {Energy Policy},
volume = {125},
pages = {330-337},
year = {2019},
issn = {0301-4215},
doi = {https://doi.org/10.1016/j.enpol.2018.11.006},
author = {Sitao Li and Sufang Zhang and Philip Andrews-Speed},
}

@article{munoz2018economic,
  title={Economic inefficiencies of cost-based electricity market designs},
  author={Munoz, Francisco D and Wogrin, Sonja and Oren, Shmuel S and Hobbs, Benjamin F},
  journal={The Energy Journal},
  volume={39},
  number={3},
  pages={51--68},
  year={2018},
  publisher={SAGE Publications Sage CA: Los Angeles, CA}
}

@book{antonopoulos2020nodal,
  title={Nodal pricing in the European internal electricity market},
  author={Antonopoulos, Georgios A and Vitiello, Silvia and Fulli, Gianluca and Masera, Marcelo and others},
  volume={30155},
  year={2020},
  publisher={Publications Office of the European Union Luxembourg}
}

@article{HERRERO201542,
title = {Electricity market-clearing prices and investment incentives: The role of pricing rules},
journal = {Energy Economics},
volume = {47},
pages = {42-51},
year = {2015},
issn = {0140-9883},
doi = {https://doi.org/10.1016/j.eneco.2014.10.024},
url = {https://www.sciencedirect.com/science/article/pii/S0140988314002709},
author = {Ignacio Herrero and Pablo Rodilla and Carlos Batlle},
}

@article{HOGAN201423,
title = {Electricity Market Design and Efficient Pricing: Applications for New England and Beyond},
journal = {The Electricity Journal},
volume = {27},
number = {7},
pages = {23-49},
year = {2014},
issn = {1040-6190},
doi = {https://doi.org/10.1016/j.tej.2014.07.009},
url = {https://www.sciencedirect.com/science/article/pii/S1040619014001705},
author = {William W. Hogan},
}

@article{FACCHINI2019110,
title = {Changes to Gate Closure and its impact on wholesale electricity prices: The case of the UK},
journal = {Energy Policy},
volume = {125},
pages = {110-121},
year = {2019},
issn = {0301-4215},
doi = {https://doi.org/10.1016/j.enpol.2018.10.047},
url = {https://www.sciencedirect.com/science/article/pii/S0301421518306992},
author = {Angelo Facchini and Alessandro Rubino and Guido Caldarelli and Giuseppe {Di Liddo}},
}

@article{cramton2017electricity,
  title={Electricity market design},
  author={Cramton, Peter},
  journal={Oxford Review of Economic Policy},
  volume={33},
  number={4},
  pages={589--612},
  year={2017},
  publisher={JSTOR}
}

@misc{data,
  author       = {{Korea MOIS}},
  title        = {Public Data Portal},
  howpublished = {\url{https://www.data.go.kr/index.do0}},
  note         = {Accessed: 25 August 2025},
  year         = {2025}
}

@misc{EIA,
  author       = {{EIA}},
  title        = {U.S Energy Information Administration},
  howpublished = {\url{https://www.eia.gov/}},
  note         = {Accessed: 27 August 2025},
  year         = {2025}
}

@article{SINSEL20202271,
title = {Challenges and solution technologies for the integration of variable renewable energy sources—a review},
journal = {Renewable Energy},
volume = {145},
pages = {2271-2285},
year = {2020},
issn = {0960-1481},
doi = {https://doi.org/10.1016/j.renene.2019.06.147},
author = {Simon R. Sinsel and Rhea L. Riemke and Volker H. Hoffmann},
}

@ARTICLE{9579026,
  author={Holttinen, Hannele and Groom, Andrew and Kennedy, Eoin and Woodfin, Dan and Barroso, Luiz and Orths, Antje and Ogimoto, Kazuhiro and Wang, Caixia and Moreno, Rodrigo and Parks, Keith and Ackermann, Thomas},
  journal={IEEE Power and Energy Magazine}, 
  title={Variable Renewable Energy Integration: Status Around the World}, 
  year={2021},
  volume={19},
  number={6},
  pages={86-96},
  doi={10.1109/MPE.2021.3104156}
}

@article{BROWN2025107484,
title = {Electricity market design with increasing renewable generation: Lessons from Alberta},
journal = {The Electricity Journal},
pages = {107484},
year = {2025},
issn = {1040-6190},
doi = {https://doi.org/10.1016/j.tej.2025.107484},
author = {David P. Brown and Derek E.H. Olmstead and Blake Shaffer},
}

@book{schweppe2013spot,
  title={Spot pricing of electricity},
  author={Schweppe, Fred C and Caramanis, Michael C and Tabors, Richard D and Bohn, Roger E},
  year={2013},
  publisher={Springer Science \& Business Media}
}

@article{cadwalader2010extended,
  title={Extended LMP and financial transmission rights},
  author={Cadwalader, Michael and Gribik, Paul and Hogan, William and Pope, Susan},
  journal={Harvard Univ., Cambridge, MA, USA, working paper},
  year={2010}
}

@article{hogan1992contract,
  title={Contract networks for electric power transmission},
  author={Hogan, William W},
  journal={Journal of regulatory economics},
  volume={4},
  number={3},
  pages={211--242},
  year={1992},
  publisher={Springer}
}

@misc{pope2014price,
  title={Price formation in ISOs and RTOs, principles and improvements. October 2014. FTI Consulting},
  author={Pope, SL},
  year={2014}
}

@misc{KPX,
  author       = {{KPX}},
  title        = {Demand Forecast for Day-ahead Unit-commitment},
  howpublished = {\url{https://www.kpx.or.kr/}},
  note         = {Accessed: 28 August 2025},
  year         = {2025} 
}

@book{akcura2024global,
  title={Global Evolution of Power Market Designs},
  author={Akcura, Elcin and Mutambatsere, Emelly},
  journal={Policy Research working paper},
  year={2024},
  publisher={World Bank}
}

@article{LESLIE2025107489,
title = {How new (and old) power system operating constraints map to Australia's wholesale electricity market model},
journal = {The Electricity Journal},
pages = {107489},
year = {2025},
issn = {1040-6190},
doi = {https://doi.org/10.1016/j.tej.2025.107489},
author = {Gordon W. Leslie and Farhad Billimoria},
}

@article{solar2024wind,
  title={Integrating Solar and Wind: Global experience and emerging challenges},
  author={IEA},
  journal={URL: https://iea. blob. core. windows. net/assets/4e495603-7d8b-4f8b-8b60-896a5936a31d/IntegratingSolarandWind. pdf},
  year={2024}
}

@article{korea2025,
  title={Energy Policy Review Korea 2025},
  author={IEA},
  journal={URL: https://iea.blob.core.windows.net/assets/21b0a0d2-6b9a-435d-b591-c65fb76399d6/Korea2025.pdf},
  year={2025}
}

@ARTICLE{9347727,
  author={Zhang, Congyue and Dou, Xiaobo and Zhang, Zhang and Lou, Guannan and Yang, Fan and Li, Guixin},
  journal={IEEE Transactions on Power Systems}, 
  title={Inertia-Enhanced Distributed Voltage and Frequency Control of Low-Inertia Microgrids}, 
  year={2021},
  volume={36},
  number={5},
  pages={4270-4280},
  doi={10.1109/TPWRS.2021.3057078}}

@ARTICLE{Lee,
    author = {S.Lee},
    title = {Frequency stability analysis considering additional tripping of renewable generation following disturbances in the Jeju power system},
    journal = {Proceedings of the KIEE Power Engineering Society Conference},
    year = {2021}
}

@article{JUNG20231374,
title = {Evaluation of inertia resource for securing nadir frequency in HVDC interconnected system with high penetration of RES},
journal = {Energy Reports},
volume = {9},
pages = {1374-1383},
year = {2023},
note = {2022 The 3rd International Conference on Power and Electrical Engineering},
issn = {2352-4847},
doi = {https://doi.org/10.1016/j.egyr.2023.05.192},
author = {Jaeyeop Jung and Jeonghoo Park and Hwanik Lee and Byongjun Lee},
}

@article{Im2024,
author = {Im, Seunghyuk and Lee, Kyungsang and Lee, Byongjun},
title = {Estimation of maximum non-synchronous generation of renewable energy in the South Korea power system based on the minimum level of inertia},
journal = {IET Renewable Power Generation},
volume = {18},
number = {7},
pages = {1260-1268},
year = {2024}
}

@article{SALEEM2024110184,
title = {Assessment of frequency stability and required inertial support for power grids with high penetration of renewable energy sources},
journal = {Electric Power Systems Research},
volume = {229},
pages = {110184},
year = {2024},
issn = {0378-7796},
doi = {https://doi.org/10.1016/j.epsr.2024.110184},
author = {M.I. Saleem and S. Saha},
}

@article{__a1___2021,
title={Details and implications of reformed Korean day-ahead electricity market}, 
volume={70}, 
DOI={10.5370/KIEE.2021.70.7.969},
number={7}, 
journal={The transactions of The Korean Institute of Electrical Engineers}, 
publisher={The Korean Institute of Electrical Engineers}, 
author={Ok, KiYoul and Lee, SungWoo and Park, MinSu and Ju, Anziin and Cho, SungBong},
year={2021},
}

@techreport{park2023korean,
  title={Korean Power System Challenges and Opportunities, Priorities for Swift and Successful Clean Energy Deployment at Scale},
  institution = {Lawrence Berkeley National Laboratory},
  author={Park, Won Young and Khanna, Nina and Kim, James Hyungkwan and Shiraishi, Kenji and Abhyankar, Nikit and Paliwal, Umed and Lin, Jiang and Phadke, Amol and Moon, Hee Seung and Song, Yong Hyun and others},
  year={2023}
}

@article{HOHL2023113503,
title = {Intraday markets, wind integration and uplift payments in a regional U.S. power system},
journal = {Energy Policy},
volume = {175},
pages = {113503},
year = {2023},
issn = {0301-4215},
doi = {https://doi.org/10.1016/j.enpol.2023.113503},
author = {Cody Hohl and Chiara {Lo Prete} and Ashish Radhakrishnan and Mort Webster},
}

@techreport{KOTRA2024,
  title        = {In-depth Analysis of the Australian BESS Industry for Accelerating Carbon Neutrality (Korean)},
  author       = {KOTRA},
  institution  = {Korea Trade-Investment Promotion Agency (KOTRA)},
  year         = {2024},
  url          = {https://dl.kotra.or.kr/pyxis-api/2/digital-files/918aeab8-d289-4235-a021-8d7f253b8369},
}

@misc{CAISO2026,
  author       = {CAISO},
  title        = {Key Statistics},
  year         = {2026},
  month        = jan,
  url          = {https://www.caiso.com/library/key-statistics},
}

@misc{KPX2025,
  author       = {Korea Power Exchange},
  title        = {Generation Capacity Status in Korea, 2024},
  year         = {2025},
  month        = jul,
  url          = {https://www.kpx.or.kr},
  note         = {Accessed: 2026-03-13}
}

@misc{KPX2024,
  author       = {Korea Power Exchange},
  title        = {Power System Operation Statistics: December 2024},
  year         = {2025},
  month        = feb,
  url          = {https://www.kpx.or.kr/board.es?mid=a10102000000&bid=0159&act=view&list_no=74511},
  note         = {Accessed: 2026-03-13}
}

@misc{KPX2025_DR,
  author       = {{Korea Power Exchange}},
  title        = {Demand Response Market Status: November 2025},
  year         = {2026},
  month        = jan,
  url          = {https://new.kpx.or.kr/board.es?mid=a10102000000&bid=0088&act=view&list_no=76610},
  note         = {Accessed: 2026-03-13}
}

@article{maliszewski1996available,
  title={Available transfer capability definitions and determination},
  author={Maliszewski, RM and Rozier, G and Cummings, R},
  journal={Princeton, NJ: North American Electric Reliability Council},
  year={1996}
}

@article{hirst2001real,
  title={Real-time balancing operations and markets: key to competitive wholesale electricity markets},
  author={Hirst, Eric},
  journal={Edison Electric Institute},
  year={2001}
}

@article{HOGAN201633,
title = {Virtual bidding and electricity market design},
journal = {The Electricity Journal},
volume = {29},
number = {5},
pages = {33-47},
year = {2016},
issn = {1040-6190},
doi = {https://doi.org/10.1016/j.tej.2016.05.009},
url = {https://www.sciencedirect.com/science/article/pii/S104061901630063X},
author = {William W. Hogan},
}

@article{hortaccsu2008understanding,
  title={Understanding strategic bidding in multi-unit auctions: a case study of the Texas electricity spot market},
  author={Horta{\c{c}}su, Ali and Puller, Steven L},
  journal={The RAND Journal of Economics},
  volume={39},
  number={1},
  pages={86--114},
  year={2008},
  publisher={Wiley Online Library}
}

\end{document}